\documentstyle[aps,prd,epsfig,preprint]{revtex}
\def\fun#1#2{\lower3.6pt
\vbox{\baselineskip0pt\lineskip.9pt
\ialign{$\mathsurround=0pt#1\hfill##\hfil$
\crcr#2\crcr\sim\crcr}}}
\newcommand{\newc}{\newcommand}
\newc{\ra}{\rightarrow}
\newc{\lra}{\leftrightarrow}
\newc{\beq}{\begin{equation}}
\newc{\eeq}{\end{equation}}
\newc{\barr}{\begin{eqnarray}}
\newc{\earr}{\end{eqnarray}}
\begin{document}
\vspace{0.5in}
\title{\vskip-2.5truecm{\hfill \baselineskip 14pt
\vskip .1truecm}
\vspace{1.0cm}
\vskip 0.1truecm{
THEORETICAL DIRECTIONAL AND MODULATED RATES
FOR DIRECT SUSY DARK MATTER DETECTION
 }}
\vspace{1cm}
\author{J. D. Vergados$^{(1),(2)}$\thanks{vergados@cc.uoi.gr}}
\vspace{1.0cm}
\address{$^{(1)}${\it Theoretical Physics Division, University of Ioannina,
Ioannina, Gr 451 10, Greece.}}
\address{$^{(2)}${\it Institute of Theoretical Physics, University of Tuebingen,Tuebingen, Germany.}}
\maketitle

\vspace{.8cm}

\begin{abstract}
\baselineskip 12pt
\par
Exotic dark matter together with the vacuum energy (cosmological constant)
seem to dominate in the flat Universe.  Thus direct dark matter detection
is central to particle physics and cosmology. Supersymmetry
provides a natural dark matter candidate, the lightest
supersymmetric particle (LSP).
 It is possible to obtain detectable rates, but realystically they are
 expected to be much lower than the present experimental goals.
So one should exploit
 two characteristic signatures of the reaction, namely
the modulation effect and the correlation with the sun's motion in
 directional experiments.
 In standard non directional experiments the modulation is small,
less than two per cent. In the case of the directional experiments, the main
 subject of this
paper, we find two novel features, which are essentially
independent of the SUSY model employed, namely: 1) The
forward-backward asymmetry, with respect to the sun's direction of
motion, is very large and 2) The modulation observed in a plane
 perpendicular to the sun's motion can be  higher than 20 per cent and
 is direction dependent.
\end{abstract}

\newpage
\pagestyle{plain}
\setcounter{page}{1}
\baselineskip 20pt
\date{\today}
\section{Introduction}
The combined MAXIMA-1 \cite{MAXIMA-1}, BOOMERANG \cite{BOOMERANG},
DASI \cite{DASI}
and COBE/DMR Cosmic Microwave Background (CMB) observations
\cite{COBE} imply that the Universe is flat \cite{flat01},
 $\Omega=1.11\pm0.07$, while the baryonic component is very small
$\Omega_b h^2=0.032^{+0.009}_{~0.008}$.
Furthermore exotic dark matter has become necessary
in order to close the Universe. In fact about a decade ago
 the COBE data ~\cite{COBE} suggested that CDM (Cold Dark Matter)
dominates the Universe, $\Omega_{CDM}$ being at least $60\%$ ~\cite {GAW}.
 Subsequent evidence from two different teams,
the High-z Supernova Search Team \cite {HSST} and the
Supernova Cosmology Project  ~\cite {SPF} $^,$~\cite {SCP}
changed this view suggesting that the Universe may be dominated by
the  cosmological constant $\Lambda$ or dark energy. In other words one
roughly finds a baryonic component $\Omega_B=0.1$ along with the exotic
components $\Omega _{CDM}= 0.3$ and $\Omega _{\Lambda}= 0.6$.
In  a more detailed recent $\Lambda CDM$
 analysis by Primack \cite{Primack} (see also the
analysis by Einasto \cite{Eina01}) we find
$h=0.72\pm0.08$, $\Omega_b=0.040\pm0.002$, and
 $\Omega _m=\Omega_{CDM}=0.33\pm0.035$ (from
cluster baryons etc), $0.4 |\pm0.2$ (from cluster evolution) and $0.34 \pm 0.1$
(from $Ly\alpha$ forest $P(k)$), while
$\Omega_{HDM}\le 0.05$ and $\Omega_{\Lambda}=0.73\pm0.08$. In other words
$\Omega_{m} \approx 3/4~\Omega_{\Lambda}$.
Since the non exotic component cannot exceed $40\%$ of the CDM
~\cite {Benne}, there is room for the exotic WIMP's
(Weakly  Interacting Massive Particles).
  In fact the DAMA experiment ~\cite {BERNA2}
has claimed the observation of one signal in direct detection of a WIMP, which
with better statistics has subsequently been interpreted as a modulation signal
~\cite{BERNA1}.

 Supersymmetry naturally provides candidates for the dark matter constituents
 \cite{GOODWIT}-\cite{ELLROSZ}.
 In the most favored scenario of supersymmetry the
LSP can be simply described as a Majorana fermion, a linear
combination of the neutral components of the gauginos and Higgsinos
 \cite{GOODWIT}-\cite{ref2}.  The main ingredients of the SUSY input are summarized in the next
 section. The essential features of the LSP-nucleus cross-section are discussed
 in sect. 3. The basic formulas for the event rates, in the case of Gaussian
 models as well as those due caustic rings, are given in sect. 4. The results
 obtained
both for the non directional as well as the directional experiments
 are discussed in sect. 5. Finally in sect. 6 we present our conclusions.

\section{The Essential Theoretical Ingredients  of Direct Detection.}

 It is well known that there exists indirect evidence for a halo of dark matter from the
 observed rotational curves. It is, however, essential to directly
detect \cite{GOODWIT}-\cite{KVprd}
 such matter, since this, among other things, will also
 unravel the nature of the constituents of dark matter. The
 possibility of detection depends on the nature of such
 constituents. Here we will assume that such a constituent is the
 LSP.
  Since this particle is expected to be very massive, $m_{\chi} \geq 30 GeV$, and
extremely non relativistic with average kinetic energy
$T \approx 50KeV (m_{\chi}/ 100 GeV)$, it can be directly detected
~\cite{GOODWIT}-\cite{KVprd} mainly via the recoiling of a nucleus (A,Z) in
elastic scattering. The event rate for such a process can be
computed from the following ingredients:

1) An effective Lagrangian at the elementary particle
(quark) level obtained in the framework of supersymmetry as described
, e.g., in Refs~\cite{ref2,JDV96}.

2) A well defined procedure for transforming the amplitude
obtained from the previous effective Lagrangian from the quark to
the nucleon level, i.e. a quark model for the nucleon. This step
is not trivial since the obtained results depend crucially on the
content of the nucleon in quarks other than u and d. This is
particularly true for the scalar couplings, which are proportional
to the quark masses~\cite{drees}$-$\cite{Chen} as well as the
isoscalar axial coupling.

3) Compute the relevant nuclear matrix elements
\cite{Ress}$-$\cite{DIVA00}. using as reliable as possible many
body nuclear wave functions. Fortunately in the case of the scalar
coupling, which is viewed as the most important, the situation is
a bit simpler, since  then one  needs only the nuclear form
factor.

Since the obtained rates are very low, one would like to be able
to exploit the modulation of the event rates due to the earth's
revolution around the sun \cite{DFS86,FFG88}
\cite{Verg98}$-$\cite{Verg01}. In order to accomplish this one
adopts a folding procedure, i.e one has to assume some velocity
distribution
\cite{DFS86,COLLAR92},~\cite{Verg99,Verg01},\cite{UK01}-\cite{GREEN02}
for the LSP. One also would like to exploit the signatures
expected to show up in directional experiments, by observing the
nucleus in a certain direction. Since the sun is moving with
relatively high velocity with respect to the center of the galaxy
one expects strong correlation of such observations with the
motion of the sun \cite {ref1,UKDMC}. On top of this one expects
in addition to see a more interesting pattern of modulation.

 The calculation of this cross section  has become pretty standard.
 One starts with
representative input in the restricted SUSY parameter space as
described in the literature~\cite{Gomez,ref2} (see also Arnowitt
and Dutta \cite{ARNDU}). We have adopted a similar procedure,
which has previously described \cite{Gomez} and will not be
repeated here.
 In the above procedure the most important constraints on the SUSY parameter space come from
 \cite{ARNDU,Gomez}:\\
1.) The LSP relic abundance which must satisfy the cosmological
constrain:
\begin{equation}
0.09 \le \Omega_{LSP} h^2 \le 0.22
\label{eq:in2}
\end{equation}
 2.) The Higgs mass bound. The bound is obtained from the
 recent experiments CDF experiment \cite{VALLS}
 and, especially the LEP2 signal \cite{DORMAN}, \cite{LEP2}
  $m_h=115 ^{1.3}_{-0.9}~GeV$,
 in SUSY cannot unambiguously attributed to a definite mass
 eigenstate. Furthermore in the LSP-nucleus scattering both physical
scalar eigenstates contribute (the surviving pseudoscalar does not
lead to coherence). The correct prediction for the Higgs mass,
however, is essential, since it imposes important constraints on
the allowed
 parameter space \cite{BFS01},\cite{ADHSW}.

 3.) The $b \rightarrow s~\gamma$ limit and the bound on the
 anomalous magnetic moment of the muon, see, e.g., recent work
and references therein \cite{AADS00}.\\
4.) The need to will  ourselves to LSP-nucleon cross sections for
the scalar coupling, which gives detectable rates
\begin{equation}
4\times 10^{-7}~pb~ \le \sigma^{nucleon}_{scalar}
\le 2 \times 10^{-5}~pb~
\label{eq:in3}
\end{equation}
This is because above this range the direct observation of LSP
should have occurred in the experimental searches so far and below
this it is unobservable.
 We should remember, however, that the event rate does not depend only
 on the nucleon cross section, but on other parameters also, mainly
 on the LSP mass and the nucleus used in target.
The condition on the nucleon cross section imposes the most severe
constraints on the acceptable parameter space. In particular in
our model \cite{Gomez} as well as in other models
\cite{BDFS99,AADS00} it restricts $tan \beta$ to values $tan \beta
\simeq 50$.
\bigskip
\section{The LSP-Nucleus Expressions Differential Cross Section .}
\bigskip
The expressions for this cross-section are well known. We will,
however, summarize the main ingredients here for the reader's
convenience and in order to establish notation.

 To begin with the effective Lagrangian describing the LSP-nucleus cross section can
be conveniently put in a  form familiar from weak interactions:
\cite {JDV96} \beq {\it L}_{eff} = - \frac {G_F}{\sqrt 2} \{({\bar
\chi}_1 \gamma^{\lambda} \gamma_5 \chi_1) J_{\lambda} + ({\bar
\chi}_1
 \chi_1) J\}
 \label{eq:eg 41}
\eeq
where
\beq
  J_{\lambda} =  {\bar N} \gamma_{\lambda} (f^0_V +f^1_V \tau_3
+ f^0_A\gamma_5 + f^1_A\gamma_5 \tau_3)N~~,~~
J = {\bar N} (f^0_s +f^1_s \tau_3) N
 \label{eq:eg.42}
\eeq

We have neglected the uninteresting pseudoscalar and tensor
currents. Note that, due to the Majorana nature of the LSP,
${\bar \chi_1} \gamma^{\lambda} \chi_1 =0$ (identically).

 With the above ingredients the differential cross section can be cast in the
form \cite{ref1,Verg98,Verg99}
\begin{equation}
d\sigma (u,\upsilon)= \frac{du}{2 (\mu _r b\upsilon )^2} [(\bar{\Sigma} _{S}
                   +\bar{\Sigma} _{V}~ \frac{\upsilon^2}{c^2})~F^2(u)
                       +\bar{\Sigma} _{spin} F_{11}(u)]
\label{2.9}
\end{equation}
\begin{equation}
\bar{\Sigma} _{S} = \sigma_0 (\frac{\mu_r(A)}{\mu _r(N)})^2  \,
 \{ A^2 \, [ (f^0_S - f^1_S \frac{A-2 Z}{A})^2 \, ] \simeq \sigma^S_{p,\chi^0}
        A^2 (\frac{\mu_r(A)}{\mu _r(N)})^2
\label{2.10}
\end{equation}
\begin{equation}
\bar{\Sigma} _{spin}  =  \sigma^{spin}_{p,\chi^0}~\zeta_{spin}~~,~~
\zeta_{spin}= \frac{(\mu_r(A)/\mu _r(N))^2}{3(1+\frac{f^0_A}{f^1_A})^2}S(u)
\label{2.10a}
\end{equation}
\begin{equation}
S(u)=[(\frac{f^0_A}{f^1_A} \Omega_0(0))^2 \frac{F_{00}(u)}{F_{11}(u)}
  +  2\frac{f^0_A}{ f^1_A} \Omega_0(0) \Omega_1(0)
\frac{F_{01}(u)}{F_{11}(u)}+  \Omega_1(0))^2  \, ]
\label{2.10b2}
\end{equation}
\begin{equation}
\bar{\Sigma} _{V}  =  \sigma^V_{p,\chi^0}~\zeta_V
\label{2.10c}
\end{equation}
\begin{equation}
\zeta_V =  \frac{(\mu_r(A)/\mu _r(N))^2}{(1+\frac{f^1_V}{f^0_V})^2} A^2 \,
(1-\frac{f^1_V}{f^0_V}~\frac{A-2 Z}{A})^2 [ (\frac{\upsilon_0} {c})^2
[ 1  -\frac{1}{(2 \mu _r b)^2} \frac{2\eta +1}{(1+\eta)^2}
\frac{\langle~2u~ \rangle}{\langle~\upsilon ^2~\rangle}]
\label{2.10d}
\end{equation}
\\
$\sigma^i_{p,\chi^0}=$ proton cross-section,$i=S,spin,V$ given by:\\
$\sigma^S_{p,\chi^0}= \sigma_0 ~(f^0_S)^2~(\frac{\mu _r(N)}{m_N})^2$
 (scalar) ,
(the isovector scalar is negligible, mainly since the heavy quarks dominate
~\cite{drees}$-$\cite{Chen},  i.e. $\sigma_p^S=\sigma_n^S)$\\
$\sigma^{spin}_{p,\chi^0}=\sigma_0~~3~(f^0_A+f^1_A)^2~(\frac{\mu _r(N)}{m_N})^2$
  (spin) ,
$\sigma^{V}_{p,\chi^0}= \sigma_0~(f^0_V+f^1_V)^2~(\frac{\mu _r(N)}{m_N})^2$
(vector)   \\
where $m_N$ is the nucleon mass,
 $\eta = m_x/m_N A$, and
 $\mu_r(A)$ is the LSP-nucleus reduced mass,
 $\mu_r(N)$ is the LSP-nucleon reduced mass and
\begin{equation}
\sigma_0 = \frac{1}{2\pi} (G_F m_N)^2 \simeq 0.77 \times 10^{-38}cm^2
\label{2.7}
\end{equation}
\begin{equation}
Q=Q_{0}u~~, \qquad Q_{0} = \frac{1}{A m_{N} b^2}=4.1\times 10^4A^{-4/3}~KeV
\label{2.15}
\end{equation}
where
$Q$ is the energy transfer to the nucleus and
$F(u)$ is the nuclear form factor and $F_{11}(u)$ is the isovector spin response
factor. $S(u)$ is essentially independent of $u$. It depends crucially on
the static spin matrix elements and the ratio of the elementary isoscalar to
isovector amplitudes \cite{DIVA00}.

In the present paper we will consider both the coherent mode as well as
 the spin mode, but we will not focus on  the nucleon cross section. For the
scalar interaction we will use a form factor obtained as discussed in our
earlier work.
 We will also consider the spin contribution, which  is expected to be more
important in the case of the light targets.
 For a discussion
of the spin matrix elements we refer the reader to the literature
( see Divari et al \cite{DIVA00,JDV02}.) We only mention here that the spin
matrix elements are the largest and the most accurately calculated for the
$A=19$ system. The static spin values, however, affect the quantity
 $\bar{\Sigma}_{spin}$,
which affects the expected rate, but it is not the main subject of
the present work. What is most relevant here is the spin response
function $F_{11}(u)$. For the light nuclei it was taken from our
earlier work \cite{DIVA00} and or the $^{127}I$ target it was
obtained from the calculation of Ressel {\it et al} \cite{Ress}.

The vector contribution, which, due to the Majorana nature of the
LSP, is only a relativistic correction and, at present, below the level of
the planned experiments, can safely be neglected.

\section{Expressions for the Rates.}
 The non-directional event rate is given by:
\begin{equation}
R=R_{non-dir} =\frac{dN}{dt} =\frac{\rho (0)}{m_{\chi}} \frac{m}{A m_N}
\sigma (u,\upsilon) | {\boldmath \upsilon}|
\label{2.17}
\end{equation}
 Where
 $\rho (0) = 0.3 GeV/cm^3$ is the LSP density in our vicinity and
 m is the detector mass
The differential non-directional  rate can be written as
\begin{equation}
dR=dR_{non-dir} = \frac{\rho (0)}{m_{\chi}} \frac{m}{A m_N}
d\sigma (u,\upsilon) | {\boldmath \upsilon}|
\label{2.18}
\end{equation}
where $d\sigma(u,\upsilon )$ was given above.

 The directional differential rate \cite{ref1},\cite{Verg01} in the
direction $\hat{e}$ is given by :
\begin{equation}
dR_{dir} = \frac{\rho (0)}{m_{\chi}} \frac{m}{A m_N}
\mbox{\boldmath $\upsilon$}.\hat{e} H\mbox{\boldmath
$\upsilon$}.\hat{e})
 ~\frac{1}{2 \pi}~
d\sigma (u,\upsilon)
\label{2.20}
\end{equation}
where H the Heaviside step function. The factor of $1/2 \pi$ is
introduced, since  the differential cross section of the last equation
is the same with that entering the non-directional rate, i.e. after
an integration
over the azimuthal angle around the nuclear momentum has been performed.
In other words, crudely speaking, $1/(2 \pi)$ is the suppression factor we
 expect in the directional rate compared to the usual one. The precise
suppression factor depends, of course, on the direction of
observation. The mean value of the non-directional event rate of
Eq. (\ref {2.18}), is obtained by convoluting the above
expressions with the LSP velocity distribution $f({\bf \upsilon},
{\boldmath \upsilon}_E)$ with respect to the Earth, which moves
with velocity $\upsilon_E$ relative to the sun (see below), i.e.
is given by:
 \beq \Big<\frac{dR}{du}\Big> =\frac{\rho
(0)}{m_{\chi}} \frac{m}{A m_N} \int f(\mbox{\boldmath $\upsilon$},
\mbox{\boldmath $\upsilon$}_E)
          |\mbox{\boldmath $\upsilon$}|
                      \frac{d\sigma (u,\upsilon )}{du} d^3 \mbox{\boldmath $\upsilon$}
\label{3.10}
\eeq
 The above expression can be more conveniently written as
\beq
\Big<\frac{dR}{du}\Big> =\frac{\rho (0)}{m_{\chi}} \frac{m}{Am_N} \sqrt{\langle
\upsilon^2\rangle } {\langle \frac{d\Sigma}{du}\rangle }~,~
\langle \frac{d\Sigma}{du}\rangle =\int
           \frac{   | \mbox{\boldmath $\upsilon$}|}
{\sqrt{ \langle \upsilon^2 \rangle}} f(\mbox{\boldmath
$\upsilon$},
         \mbox{\boldmath $\upsilon$}_E)
                       \frac{d\sigma (u,\upsilon )}{du} d^3 \mbox{\boldmath $\upsilon$}
\label{3.11}
\eeq
 Now we perform the   needed integrations. First over the velocity distribution ranging
 from $a \upsilon_0 \sqrt{u}$, where  $a=[\sqrt{2} mu_r b\upsilon_0 b]^{-1}$
 and $b$ the nuclear (harmonic oscillator) length parameter, to the maximum
escape velocity $\upsilon_m$. Second over the energy transfer u
ranging from $u_{min}$ dictated by the detector energy cutoff to
$u_{max}=(\upsilon_m /(\upsilon_0 a))^2$.
 Thus we get:
 \beq
R = \bar{R}~t~
          [1 + h(a,Q_{min})cos{\alpha})]
\label{3.55a}
\eeq
 where $t$ is the total rate in the absence of
modulation, $\alpha$ is the phase of the Earth ($\alpha=0$ around
June 2nd) and $Q_{min}$ is the energy transfer cutoff imposed by
the detector. In the above expressions $\bar{R}$ is the rate
obtained in the conventional approach \cite {JDV96} by neglecting
the folding with the LSP velocity and the momentum transfer
dependence of the differential cross section, i.e. by
\beq \bar{R}
=\frac{\rho (0)}{m_{\chi}} \frac{m}{Am_N} \sqrt{\langle v^2\rangle
} [\bar{\Sigma}_{S}+ \bar{\Sigma} _{spin} + \frac{\langle \upsilon
^2 \rangle}{c^2} \bar{\Sigma} _{V}] \label{3.39b} \eeq where
$\bar{\Sigma} _{i}, i=S,V,spin$
 contain all the parameters of the
SUSY models.
 The modulation is described by the parameter $h$ .

The total  directional event rates  can be obtained in a similar
fashion by suitably modifying Eq. (\ref{3.10}) \beq
\Big<\frac{dR}{du}\Big>_{dir} =\frac{\rho (0)}{m_{\chi}}
\frac{m}{A m_N} \int f( \mbox{\boldmath $\upsilon$},
\mbox{\boldmath $\upsilon$}_E) \frac{ \mbox{\boldmath
$\upsilon$}.\hat{e} H(\mbox{\boldmath $\upsilon$}.\hat{e})}{2
\pi}~ \frac{d\sigma (u,\upsilon )}{du} d^3 {\boldmath \upsilon}
\label{3.10b} \eeq
 The role played by the velocity distribution is very clear. What
 is the proper velocity distribution to use? The best approach
 seems to be to apply the Eddington approach \cite{EDDIN}. In such
 an approach one starts from a given density as a function of
 space and one solves Poisson's equation to obtain the potential.
 Then from the functional dependence of the density on the
 potential one can construct, at least numerically, the density
 distribution in phase space, as a function of the potential $\Phi({\bf r})$
 and the velocity \cite{EDDIN},\cite{UK01,BCFS02,VEROWEN}. Evaluating
 this distribution in our vicinity yields the desired velocity
 distribution. Since this procedure can only be implemented numerically
it is very hard to incorporate in the calculation of the
directional rates. We thus follow the conventional approach and
use two phenomenological velocity distributions: i) A Gaussian
distribution, which can be either spherically symmetric or only
axially symmetric and ii) A velocity distribution prescribed by
the assumption of the late infall in the form of caustic rings.
\subsection{Gaussian Distribution}
 The Gaussian distribution, with respect to the center of the galaxy,
 is of the form:
\beq f(\mbox{\boldmath $\upsilon$}^{'},\lambda)= N(\lambda)
\frac{1}{(\upsilon _0 \sqrt{\pi})^3} ~Exp\left [-
\frac{(\lambda+1)[(\upsilon^{'}_y)^2+
(\upsilon^{'}_z)^2]+(\upsilon^{'}_x)^2}{\upsilon ^2_0} \right ]
\label{veldistr} \eeq
 where $\upsilon_0$ is equal to the velocity of the sun around the center
 of galaxy, $\lambda$ is the asymmetry parameter, assumed to be in the range
 $0\le \lambda \le 1$, and $N(\lambda)$ is a normalization
 constant, $N(0)=1$. One must of, course, transform the above distribution
 into the local coordinate system, taking into account both the
 motion of the sun as well as that of the Earth. In writing the above velocity
 distribution we have chosen a set of axes as follows:

 The z-axis is along the sun's direction of motion.

 The x-axis is the radial direction outwards.

 The y- axis is perpendicular to the galactic plane, so that the system is a
 right-hand one.

 Then
 \begin{equation}
 \mbox{\boldmath $\upsilon$}^{'}=\mbox{\boldmath $\upsilon$}+\upsilon_0 \hat{z}+\mbox{\boldmath $\upsilon$}_E~,
 ~\mbox{\boldmath $\upsilon$}_E=\upsilon_E \left[ \sin{\alpha} \hat{x}-\cos{\alpha}
 \cos{\gamma} \hat{y}+ \cos{\alpha} \sin{\gamma} \hat{z} \right]
 \label{vearth}
\end{equation}
with $\gamma \approx \pi/6$ and $\alpha$ the phase of the Earth.
After that  one must do the folding with the above velocity
distribution. The integration, however, of equation \ref{3.10b} is
quite difficult due to the presence of the Heavyside function. So
for the purpose of integration we found it convenient to go to a
coordinate system in which the polar axis is in the direction of
of observation $\hat{e}$, which in the above coordinate system is
specified by the polar angle $\Theta$ and the azymouthal angle
$\Phi$. In this new
 coordinate system polar angle specifying the velocity vector
is simply restricted to be
 $0\le \theta \le \pi$, while the azymouthal angle $\phi$ is unrestricted.
Thus the unit vectors along the new coordinate axes,
$\hat{X},\hat{Y},\hat{Z}$, are expressed in terms of the old ones
as follows:
 \beq \hat{Z}=\sin \Theta \cos \Phi \hat{x}+ \sin
\Theta \sin \Phi \hat{y}+ \cos \Theta \hat{z}. \label{3.20} \eeq
\beq \hat{X}=\cos \Theta \cos \Phi \hat{x}+ \cos \Theta \sin \Phi
\hat{y}- \sin \Theta \hat{x}. \label{3.21} \eeq \beq \hat{Y}=-\sin
\Phi \hat{x}+ \cos \Phi \hat{y}. \label{3.22} \eeq Thus the LSP
velocity is expressed in the new coordinate system as: \beq
\upsilon_x=\sin \Theta \cos \Phi \upsilon_X+ \sin \Theta \sin \Phi
\upsilon_Y+ \cos \Theta \upsilon_Z, \label{3.23a} \eeq \beq
\upsilon_y=\cos \Theta \cos \Phi \upsilon_X+ \cos \Theta \sin \Phi
\upsilon_Y- \sin \Theta \upsilon_Z, \label{3.24h} \eeq \beq
\upsilon_z=-\sin \Phi \upsilon_X+ \cos \Phi \upsilon_Y,
\label{3.25} \eeq with $\upsilon_X=\upsilon \sin \theta \cos
\phi$, $\upsilon_Y=\upsilon \sin \theta \sin \phi$,
$\upsilon_Z=\upsilon \cos \theta $. It is thus straightforward to
go to polar coordinates in velocity space and get: \beq
\Big<\frac{dR}{du}\Big>_{dir} =\frac{\rho (0)}{m_{\chi}}
\frac{m}{A m_N} \int^{\upsilon_m}_{a \upsilon_0 \sqrt{u}}
\upsilon^3 d \upsilon \int^1_0 d \xi \int^{2 \pi}_0 d \phi
\frac{\tilde{f}({\Theta,\Phi, \upsilon , \upsilon_E , \xi, \phi )}
} {2 \pi}
                      \frac{d\sigma (u,\upsilon )}{du}
\label{3.26}
\eeq
 with $\xi=cos{\theta}$. Now the orientation parameters $\Theta$ and
$\Phi$ appear explicitly in the distribution function and not
implicitly via the limits of integration. The function $\tilde{f}$
can be obtained from the velocity distribution, but it will not be
explicitly shown here. Thus we obtain:
 \beq R_{dir}  =   \bar{R}
(t_{dir}/2 \pi) \,
           \left [1 + h_m cos{(\alpha-\alpha_m~\pi)} \right ]
\label{4.56}
\eeq
where the quantity $t_{dir}$ provides the non modulated amplitude,
 while $h_m$ describes the modulation amplitude and $\alpha_m$ is the
  shift in phase (in units of $\pi$), giving the  phase
of the Earth in which the maximum modulation occurs. Clearly these
parameters are functions of  $\Theta$ and $\Phi$ as well as the
parameters $a$ and $Q_{min}$. The dependence on $a$ comes from
 the nuclear form factor the folding with LSP velocity.
 The  other SUSY parameters have all been absorbed in $\bar{R}$.

 Instead of $t_{dir}$ itself it is more convenient
to present the reduction factor of the non modulated directional rate compared
to the usual non-directional one, i.e.
\beq
f_{red}=\frac{R_{dir}}{R}=t_{dir}/(2 \pi~t)= \kappa/(2 \pi)
\label{kappa.eq}
\eeq
  It turns out that the parameter $\kappa$, being the
ratio of two rates, is less dependent on the parameters of the
theory.
The directional rate can be cast in an even better form as follows:
 \beq R_{dir}  =   \bar{R}~t
\frac{\kappa}{2 \pi} \left [1 + h_m cos{(\alpha-\alpha_m~\pi)} \right]
\label{4.57}
\eeq
Another quantity, which may be of experimental interest is the
asymmetry $As=|R_{dir}(-)-R_{dir}(+)|/(R_{dir}(-)+R_{dir}(+))$ in
some given direction (+) and its opposite (-). The most relevant direction
 for the asymmetry is that of the velocity of the sun.  $As$ is almost
 independent of all other parameters except the direction of observation and
 the velocity distribution.
 The directional rates exhibit interesting pattern of modulation. From  the
functions $h_m(a,Q_{min})$ and $\alpha_m$ obtained in the present work,
 it is trivial to plot the the
expression (\ref{4.57}) as a function of the phase of the earth
 $\alpha$.  Hence, this will not be done here.
\subsection{Caustic Rings}
 One would like to examine a non isothermal model to see what
 effect, if any, may have on the directional rate.
 The model of  caustic rings proposed
 by Sikivie\cite{SIKIVIE} comes to mind
 \cite{GEL01},\cite{Verg01}. Admittedly, however, this scenario
 for dark matter distribution in the galaxy is not broadly
 accepted \cite{MHWS}, since it has not been supported
 by galaxy evaluation simulations.
 . From previous work \cite{Verg01}
 we take the data for caustic rings needed for our purpose , which, for the
  reader's convenience, we have included in Table \ref{table.caus}.

The above expressions simplify significantly in the case of caustic rings
whereby the velocity distribution is discreet.
Thus Eq. (\ref{3.11}) takes the form
\beq
\langle \frac{d\Sigma}{du} \rangle  = \frac{2 \bar{\rho}}{\rho(0)}~a^2
                          [\bar {\Sigma} _{S} \bar {F}_0(u) +
        \frac{\langle \upsilon ^2 \rangle}{c^2}\bar {\Sigma} _{V} \bar {F}_1(u)
                          +\bar {\Sigma} _{spin} \bar {F}_{spin}(u)]
\label{3.23}
\eeq
We remind the reader that $\bar{\rho}$ was obtain for each type of flow
(+ or -), which explains  the factor of two. in the Sikivie model
\cite {SIKIVIE} we have $(2\bar{\rho}/\rho(0)=1.25$.
The quantities
$\bar{F}_0,\bar{F}_1,\bar{F}_{spin}$
are obtained from the corresponding form factors via the equations
\beq
\bar{F}_{k}(u) = F^2(u)\bar{\Psi}_k(u) \frac{(1+k) }{2k+1}~~,~~ k = 0,1
\label{3.24}
\eeq
\beq
\bar{F}_{spin}(u) = F_{11}(u)\bar{\Psi}_0(u)
\label{3.24a}
\eeq
with
\beq
\tilde{\Psi}_k(u) = \sqrt{\frac{2}{3}}~\sum_{n=1}^N~ \bar{\rho}_n
                   y_{an}^{2k-1)}\Theta(\frac{y_{an}^2}{a^2}-u)
\label{3.26a}
\eeq
with
\beq
y_{an} = [(y_{nz}-1-\frac{\delta}{2}~sin\gamma~cos\alpha)^2
                  +(y_{ny}+\frac{\delta}{2}~cos\gamma~cos\alpha)^2
                  + (y_{nx}-\frac{\delta}{2}~sin\alpha)^2]^{1/2}
\label{3.26b} \eeq
 with $\delta=2 (\upsilon_E/upsilon_0)$. Integrating numerically Eq. (\ref {3.23}) we
 obtain  the total
undirectional rate as a function of the phase of the Earth. Unlike our previous
work we did not make an expansion in terms of $\delta$, in order to better
deal with threshold effects. By making a Fourier decomposition of the obtained
rate to an accuracy better than
$1\%$ we find :
\beq
R =  \bar{R}\, t \, \frac{2 \bar{\rho}}{\rho(0)}
          [1 + h(a,Q_{min})cos{\alpha})]
\label{3.55}
\eeq
 In other words the modulation is again described a parameter $h$, which,
 of course, takes different numerical values compared to those of the
 Gaussian distribution.
 Note that, unlike our earlier work \cite{Verg01}, the modulation $h$ in this
work is defined with a $+$ sign. So our present value of $h$ in the case of
caustic rings is expected to have
a phase difference of $\pi$  compared to that of the usual Gaussian
 distribution.

 Similarly integrating Eq. (\ref{3.10b}) we get
\beq
\langle \frac{d\Sigma}{du} \rangle_{dir}  = \frac{2 \bar{\rho}}
                                         {\rho(0)}~\frac{a^2}{2 \pi}
                          [\bar {\Sigma} _{S} F_0(u) +
                     \frac{\langle \upsilon ^2 \rangle}{c^2}
                          \bar {\Sigma} _{V} F_1(u) +
                          \bar {\Sigma} _{spin} F_{spin}(u) ]
\label{4.39}
\eeq
The quantities $F_0,F_1,F_{spin}$ are obtained
from the equations
 \beq F_k(u) = F^2(u)\Psi_k(u)\frac{(1+k)}{2k+1}
, k = 0,1
 \label{4.40}
 \eeq
 \beq F_{spin}(u) = F_{11}(u) \Psi_0(u)
\label{4.41} \eeq
\beq
\Psi_k(u) =\sqrt[\frac{2}{3}]~\sum_{n=1}^N~ \bar{\rho}_n
                   y_{an}^{2(k-1)}~X~H(X)
\label{4.42a}
\eeq
with $H(X)$ the usual Heaviside (theta) function and $X$ given by
\beq
X= (y_{nz}-1-\frac{\delta}{2}~sin\gamma~cos\alpha) {\bf e}_z.{\bf e}
     +(y_{ny}+\frac{\delta}{2}~cos\gamma~cos\alpha){\bf e}_y.{\bf e}
      +(y_{nx}-\frac{\delta}{2}~sin\alpha){\bf e}_x.{\bf e}
\label{4.42b}
\eeq

Note that in $X$ only certain rings contribute for a given direction of
observation $\hat{e}$ (as dictated by the Heaviside function). Note further
 that the other Heaviside function for a given ring $n$ restricts the
contribution of the form factor as follows:
\beq
\chi(y_{an})=\int_{u_{min}}^{y^2_{an}/a^2}F^2(u)du
\label{4.51}
\eeq
and an analogous expression for the spin response function.

Once again we did not make an expansion in powers of delta. We Fourier
decomposed the final expression and we found that Fourier components
higher than unity ($n \ge 2$) are negligible. In other words our results
for the directional rate can be
cast in the form of Eq. (\ref{4.57}).


\section{Results and Discussion}
The three basic ingredients for the event rate for the LSP-nucleus
scattering  are the input SUSY parameters, a quark model for the
nucleon and the velocity distribution combined with the structure
of the nuclei involved.
 In the present work we will present our  results for the coherent scattering and make
comparisons with the spin contribution, whenever possible.  We
will focus our discussion on the light targets (e.g. A=19, 23, 29)
and the more popular target $^{127}I$.

We have utilized two nucleon models indicated by B and C, for their description
see our previous work \cite{JDV02}, which take into
account the presence of heavy quarks in the nucleon. We also considered
the effects on the rates of the energy cut off imposed by the detector,
 by considering two  typical cases $Q_{min}=0,~10~ KeV$.

  The experimentally determined quenching factors, see Simon
   {\it et al} and Graichen {\it et al} \cite{SIMON03,GRAICHEN}
  and references therein, defined as the ratio of a signal induced
  by nuclear recoil to that of an electron of the same energy,
  have not been included in calculating the total
  rates in the present work. These factors, which are
  functions of the energy, depend mainly on the detector material.
  For $NaI$ they have been measured \cite{SIMON03,GERBIER} down to about $10~KeV$ and
  they are constant, about $0.25$. For our calculations employing
  $Q_{min}=10~KeV$ this amounts to a reduction of the parameter $t$
  to about $25\%$ of its value presented here. We cannot estimate what the effect
   of quenching is going below $10~KeV$.
  The modulation amplitude and the reduction
  factor $\kappa$, however, being relative parameters, are not expected to be
  influenced very much by such effects.
\subsection{Isothermal Models}
\begin{figure}
\includegraphics[height=.3\textheight]{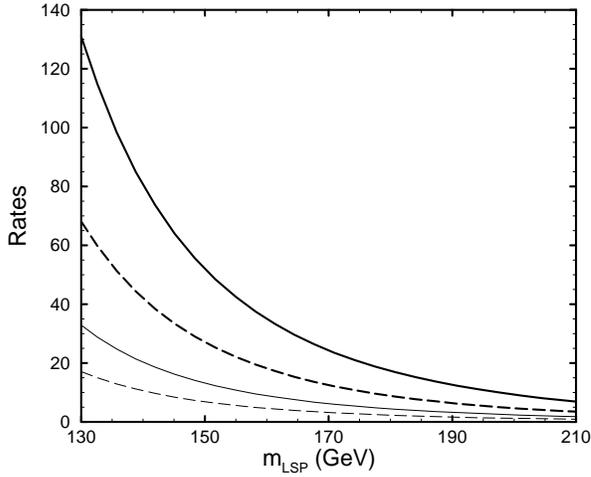}
\caption{The Total detection rate per $(kg-target)yr$ vs the LSP
mass in GeV for a typical solution of the  parameter space, as
described in our previous work (see text), in the case of
$^{127}I$. The results shown by thick lines correspond correspond
to  model B, while
 those by a thin line to model C. In the upper curve no detector cutoff was employed,
while in the lower curve we used a detector energy cutoff of
$Q_{min}=10~KeV$. Such effects introduce variations in the rates
by factors of about two. \label{rate} }
\end{figure}

For the reader's convenience we will begin by presenting results
for the unmodulated non directional event rates, $\bar{R}t$,
 of the symmetric isothermal model for a favorable SUSY parameter
 choice with large $tan(\beta)$, described previously
\cite{Gomez}, shown in Fig. \ref{rate} as a function of the LSP
mass in GeV. Clearly, depending on the SUSY parameter space and in
particular the LSP mass, the rates can change many orders of
magnitude. What we rally want to exhibit is the role of the
nucleon model employed and the effect of the energy cutoff.

\begin{figure}
\hspace*{-0.0 cm}
\includegraphics[height=.2\textheight]{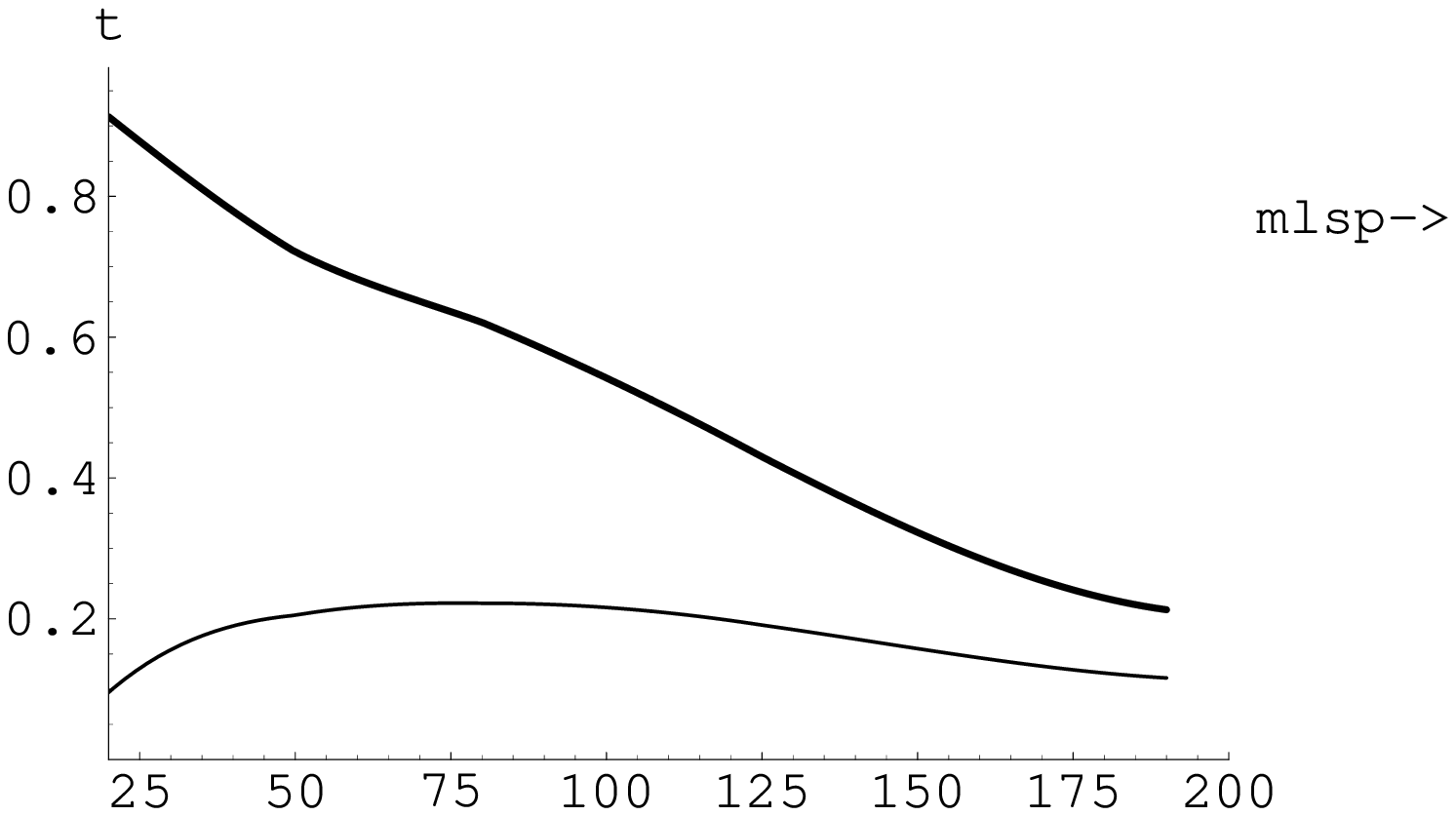}
\includegraphics[height=.2\textheight]{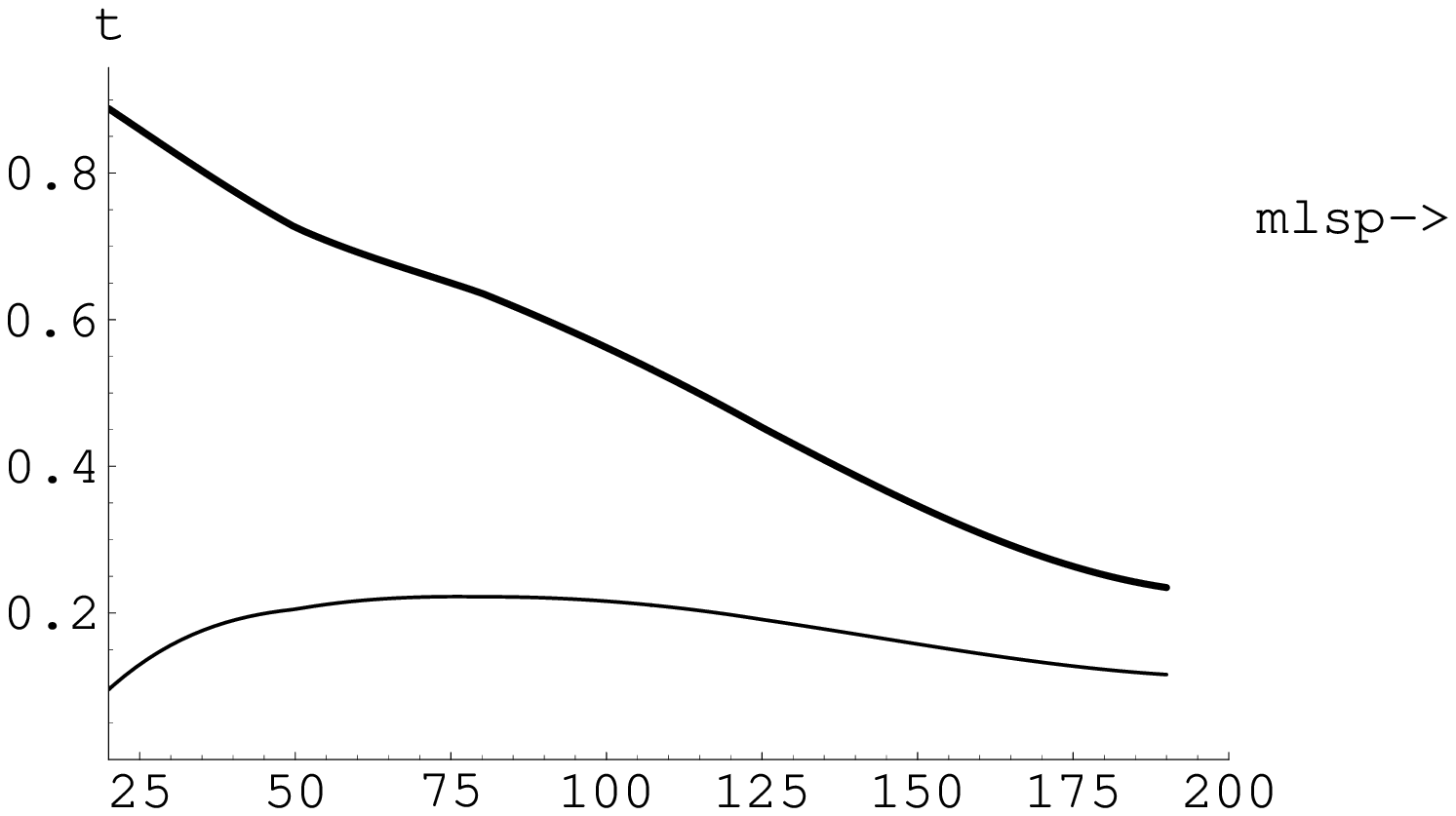}
\caption{
The dependence of the quantity $t$ on the LSP mass
 for the symmetric case ($\lambda=0$) on the left as well as for the
 maximum axial asymmetry ($\lambda=1$) on the right in the case of the
 intermediate mass target $^{127}I$.
For orientation purposes  two  detection cutoff energies are exhibited,
$Q_{min}=0$ (thick solid line) and $Q_{min}=10~keV$ (thin solid line).
 As expected $t$ decreases as the cutoff energy and/or the LSP
mass increase. We see that the parameter $\lambda$ has little
effect on the non modulated rate.
\label{fig.t}
}
\end{figure}

\begin{figure}
\hspace*{-0.0 cm}
\includegraphics[height=.2\textheight]{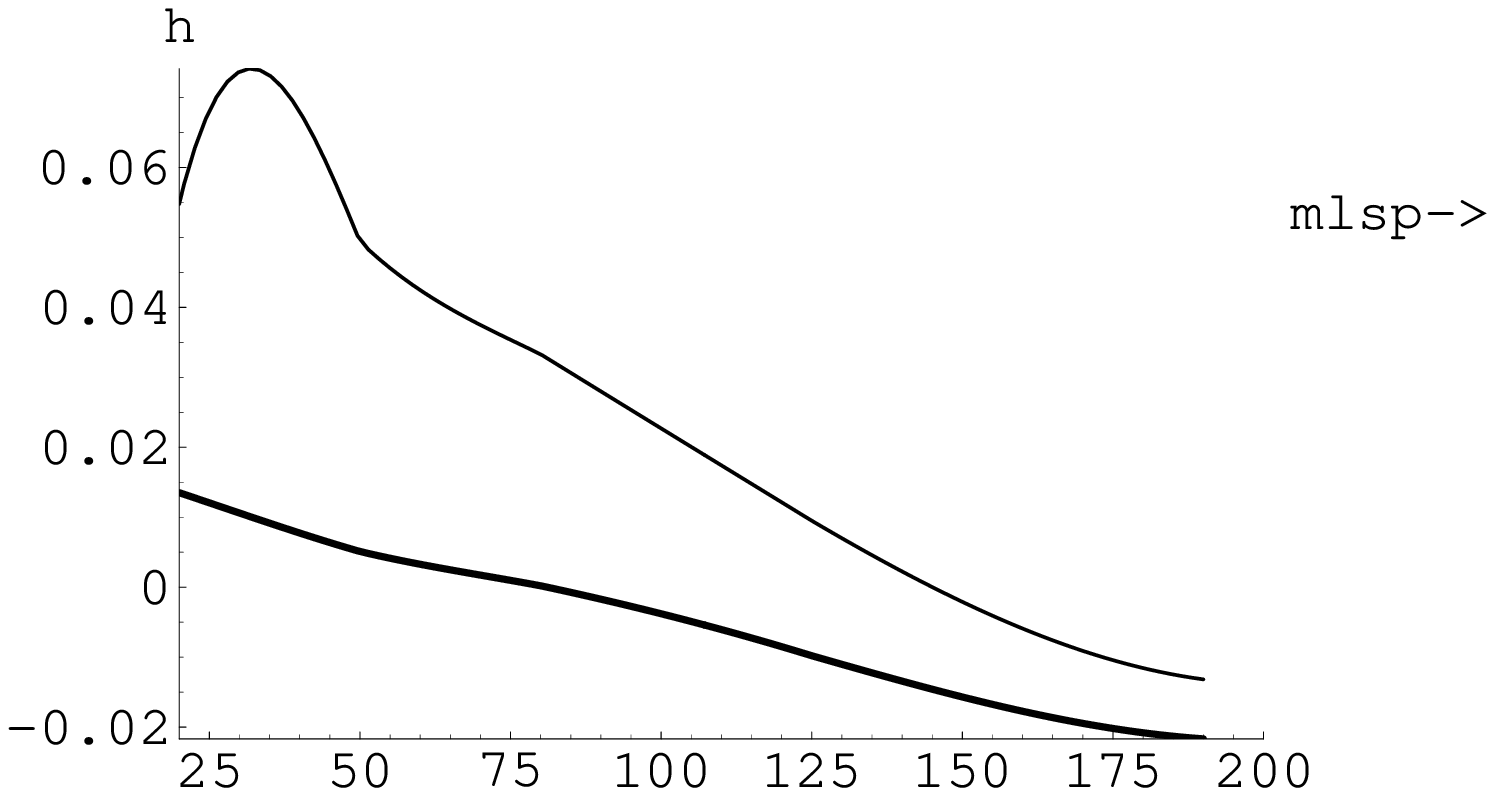}
\includegraphics[height=.2\textheight]{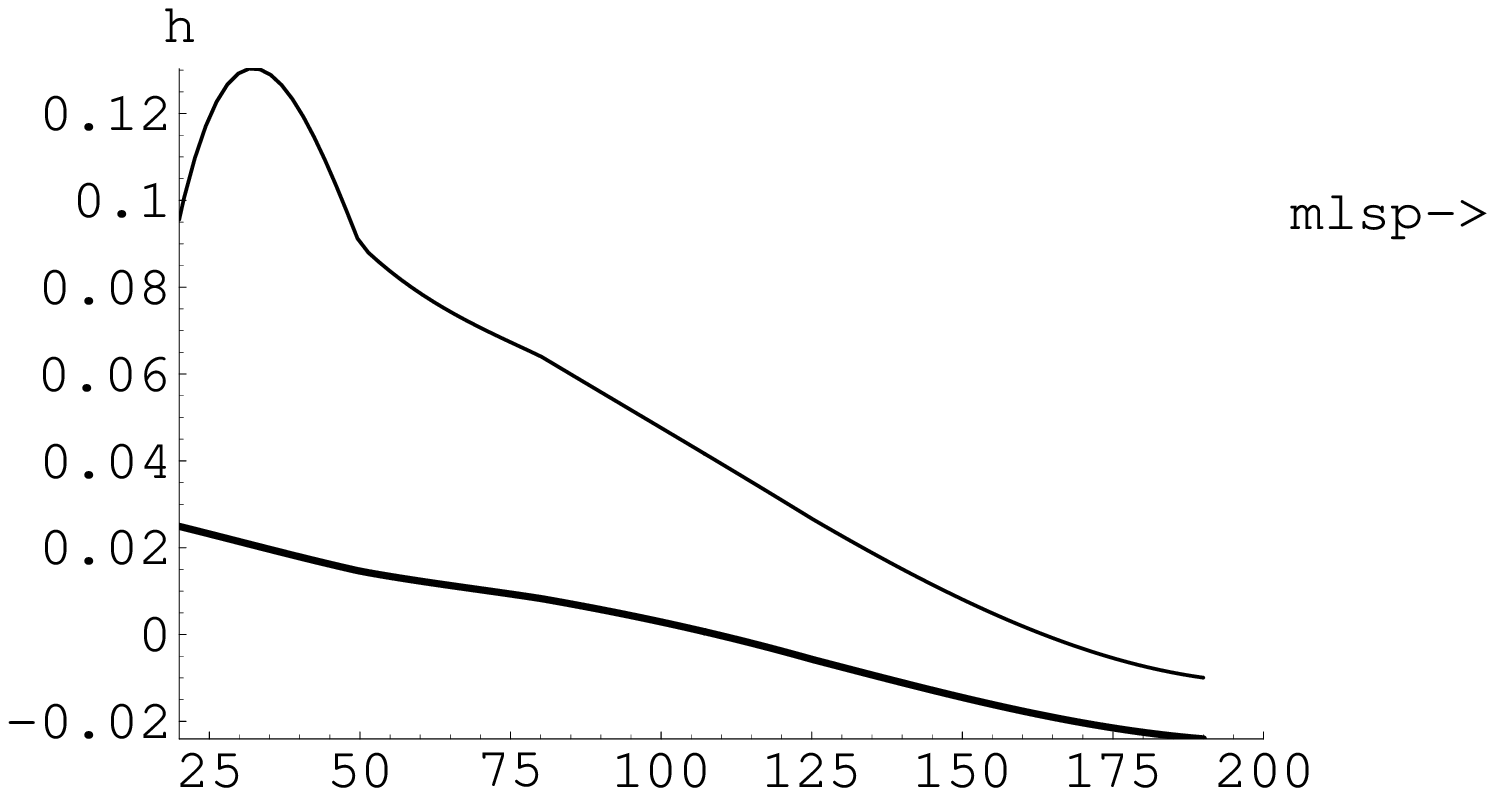}
\caption{ The same as in Fig. \ref{fig.t}
for the modulated amplitude. We see that the
modulation is small and decreases with the LSP mass.
 It even changes sign
for large LSP mass. The
 introduction of a cutoff $Q_{min}$ increases the modulation (at the expense
of the total number of counts).
 It also increases
 with the parameter $\lambda$.
\label{fig.h}
}
\end{figure}

  The two relative
parameters, i.e. the quantities $t$ and $h$, for $^{127}I$ are
shown in Figs \ref{fig.t} and \ref{fig.h} respectively in the case
of Gaussian distribution. For the light systems these quantities
are essentially constant independent of the LSP mass(the reduced
mass does not change as the LSP mass increases).They are shown in
Table \ref{table.dir}. In the case of the directional rates we
calculated the reduction factors and the asymmetry parameters as
well as the modulation amplitude
 as functions of the direction of observation, focusing our attention along
 the three axes \cite{Verg00}, i.e along $+z,-z,+y,-y,+x$ and $-x$.
In the case of the directional rates we calculated the reduction
factors and the asymmetry parameters as well as the modulation
amplitude
 as functions of the direction of observation, focusing our attention along
 the three axes \cite{Verg00}, i.e along $+z,-z,+y,-y,+x$ and $-x$.

  Since $f_{red}$ is the ratio of two parameters, its dependence
on $Q_{min}$ and the LSP mass is mild. In the case of light
nuclear systems (A=19, 23 and 29) these parameters are shown in
Table \ref{table.dir}. Note that in the favorable direction $-z$
(opposite to the velocity of the sun) the modulation is about a
factor of two bigger than in the non-directional case, but it is
still quite small ($h=0.06$). The reduction factor is
$\kappa=0.7$. In
 the sun's direction of motion the process is unobservable. In the plane
perpendicular in the sun's velocity the rate is further reduced by
a factor of about three.


 The modulation shows a very interesting pattern.
If the observation is done in the
direction opposite to the sun's direction of motion, the modulation amplitude
$h_m$ behaves in the same way as the non directional one, namely $h$. It is
more instructive to consider directions of observation in the plane
perpendicular to the sun's direction of motion ($\Theta=\pi/2$). We see from
 Table \ref{table.dir} that now the modulation is quite a bit larger, giving
rise to a difference between the maximum and the minimum rates of about
$50\%$.

For heavier nuclei the pattern slightly changes. Our results are
presented in plots, in which we adopted the following convention:
The thick solid line corresponds to $+z$ ($\Theta=0$), while the
thin line to $-z$ ($\Theta=\pi$. In the case of $\Theta=\pi/2$ we
encounter 4 cases. The intermediate thickness line corresponds to
the
 $\pm x$, the dotted line to $+y$ and the dashed line to $-y$. In some
cases two or more lines may coincide. In the case of $\kappa$ one
can distinguish only the curves corresponding to the three
$\Theta$ values.

\begin{figure}
\includegraphics[height=.3\textheight]{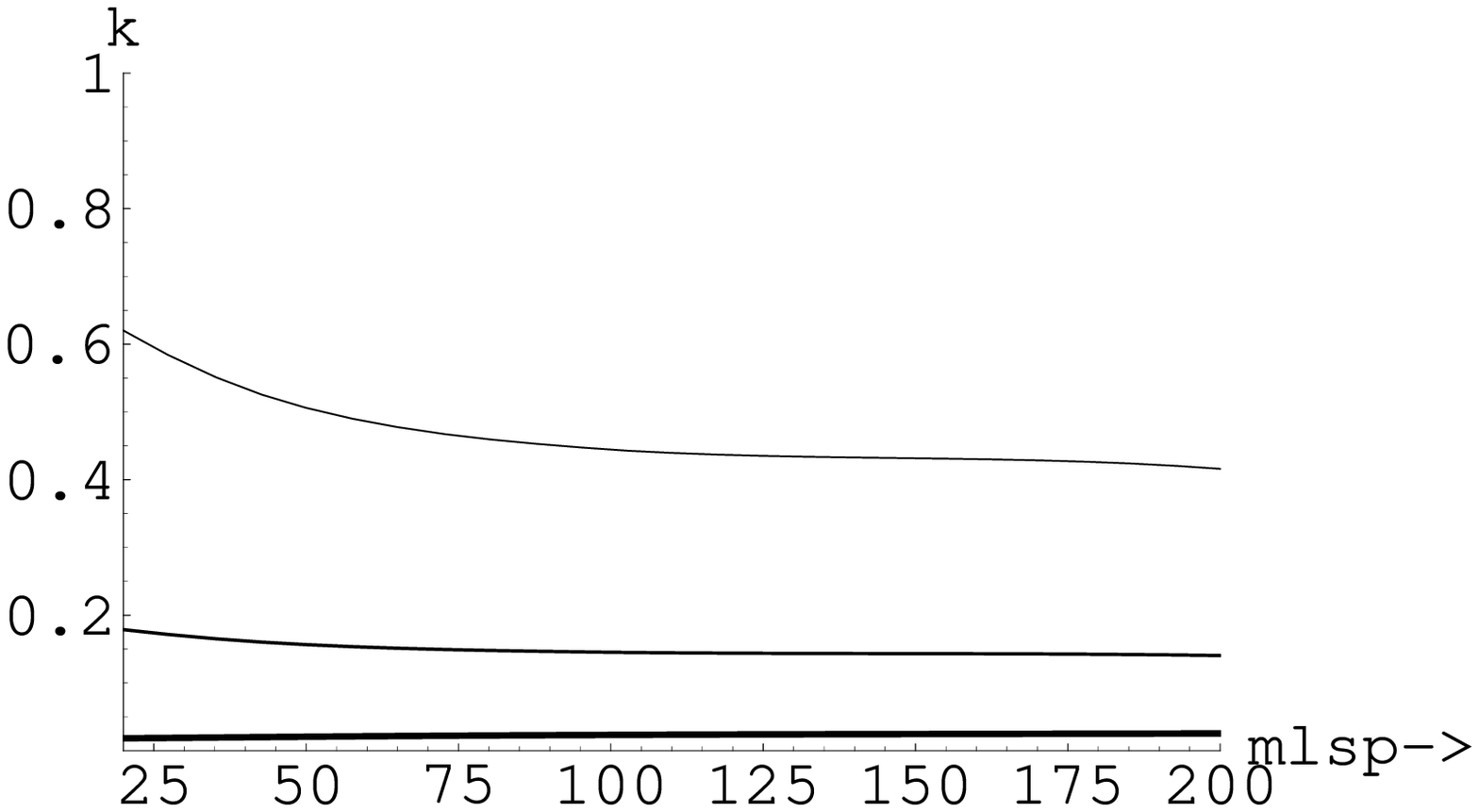}
\includegraphics[height=.3\textheight]{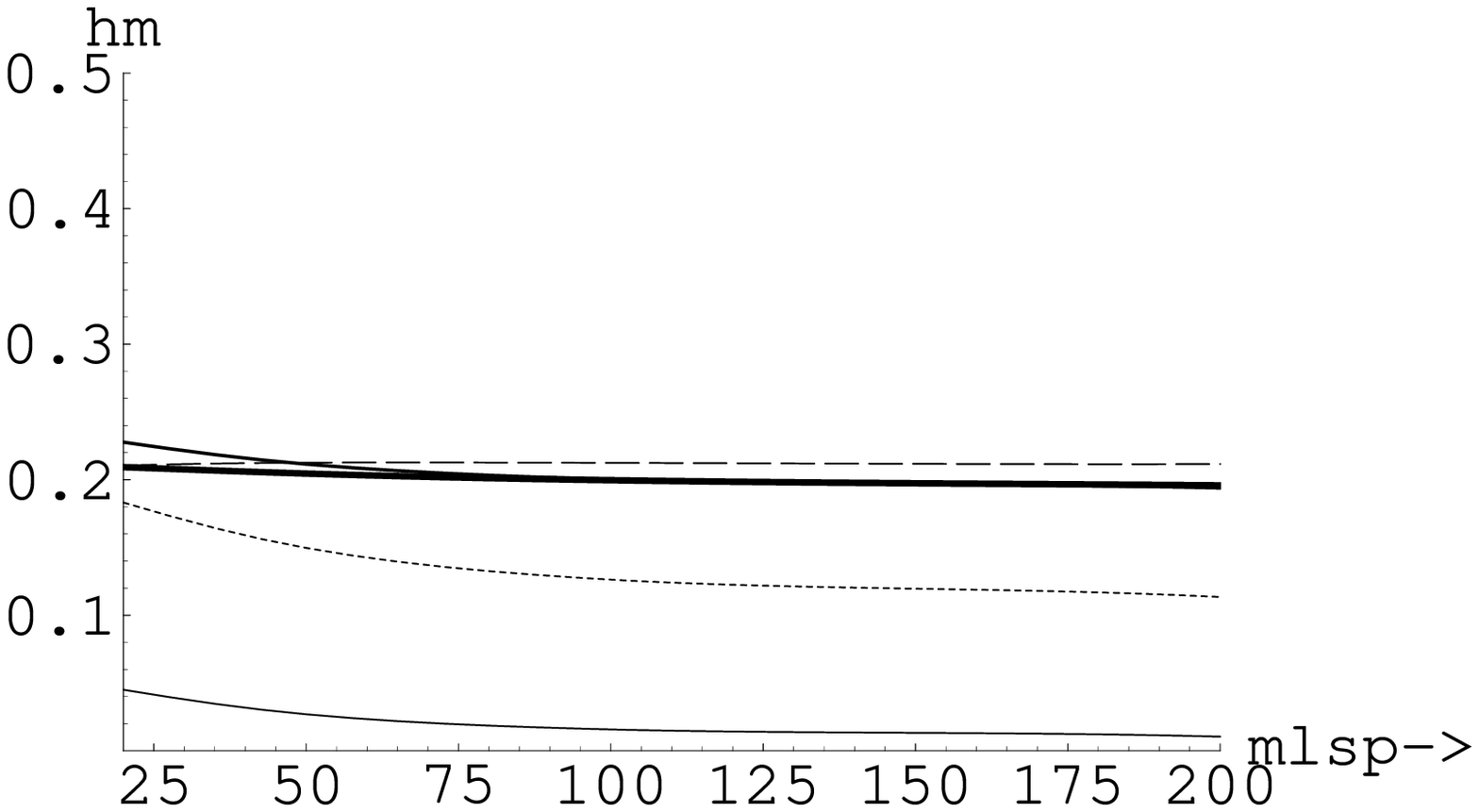}
\caption{ On the left figure one sees the quantity $\kappa$ and on
the right the quantity $h_m$ for $\lambda=0$ and $Q_{min}=0$. The
results are almost identical for the coherent and the spin modes.
They change very little for $Q_{min}=10~KeV$. For the
identification of the curves see text. Note that in the case of
the modulation the curves corresponding to $+z$ and $\pm y$
coincide. The large modulation seen in the $+z$ direction is
essentially useless since the event rate is very small.
\label{kaph0.0} }
\end{figure}

\begin{figure}
\includegraphics[height=.3\textheight]{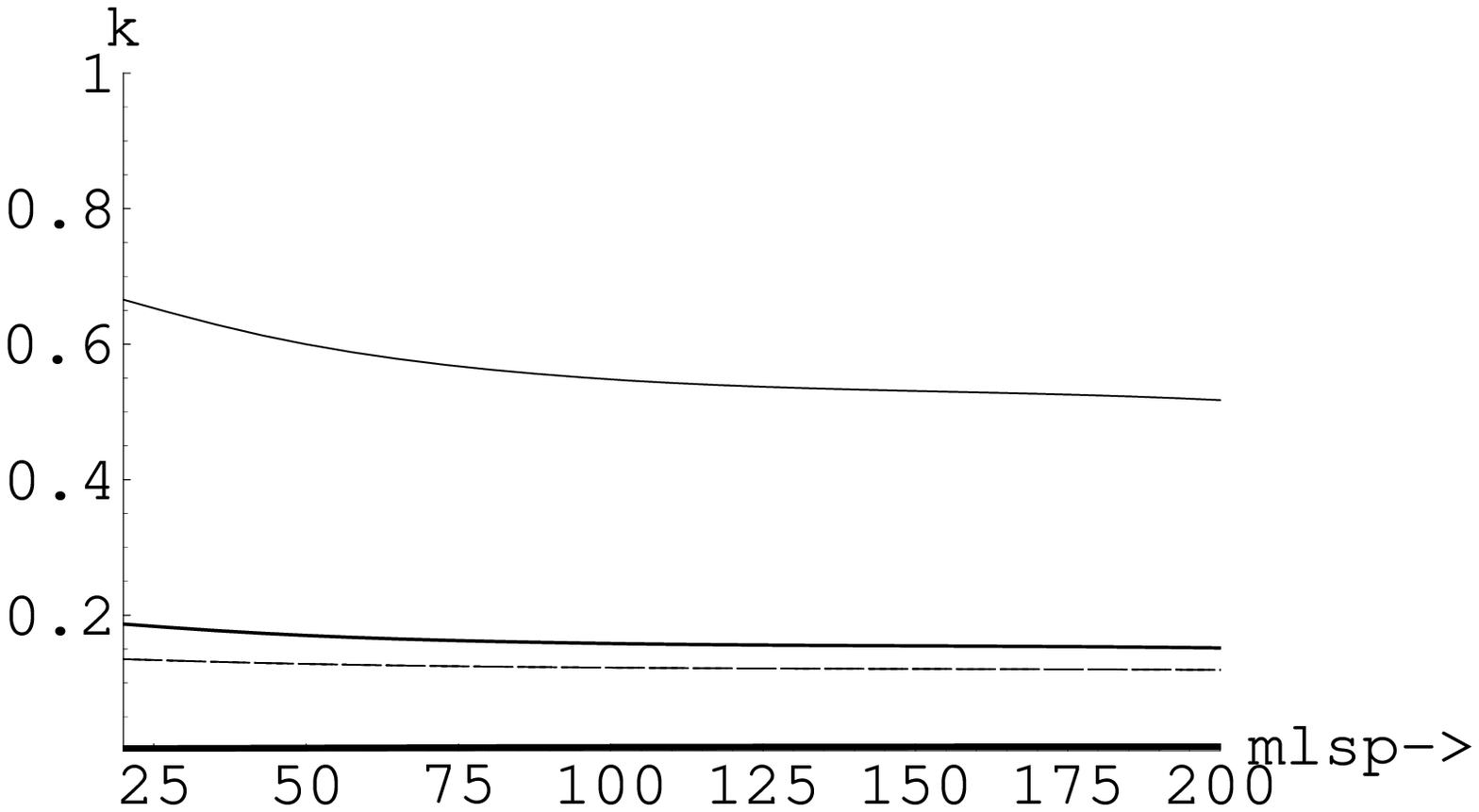}
\includegraphics[height=.3\textheight]{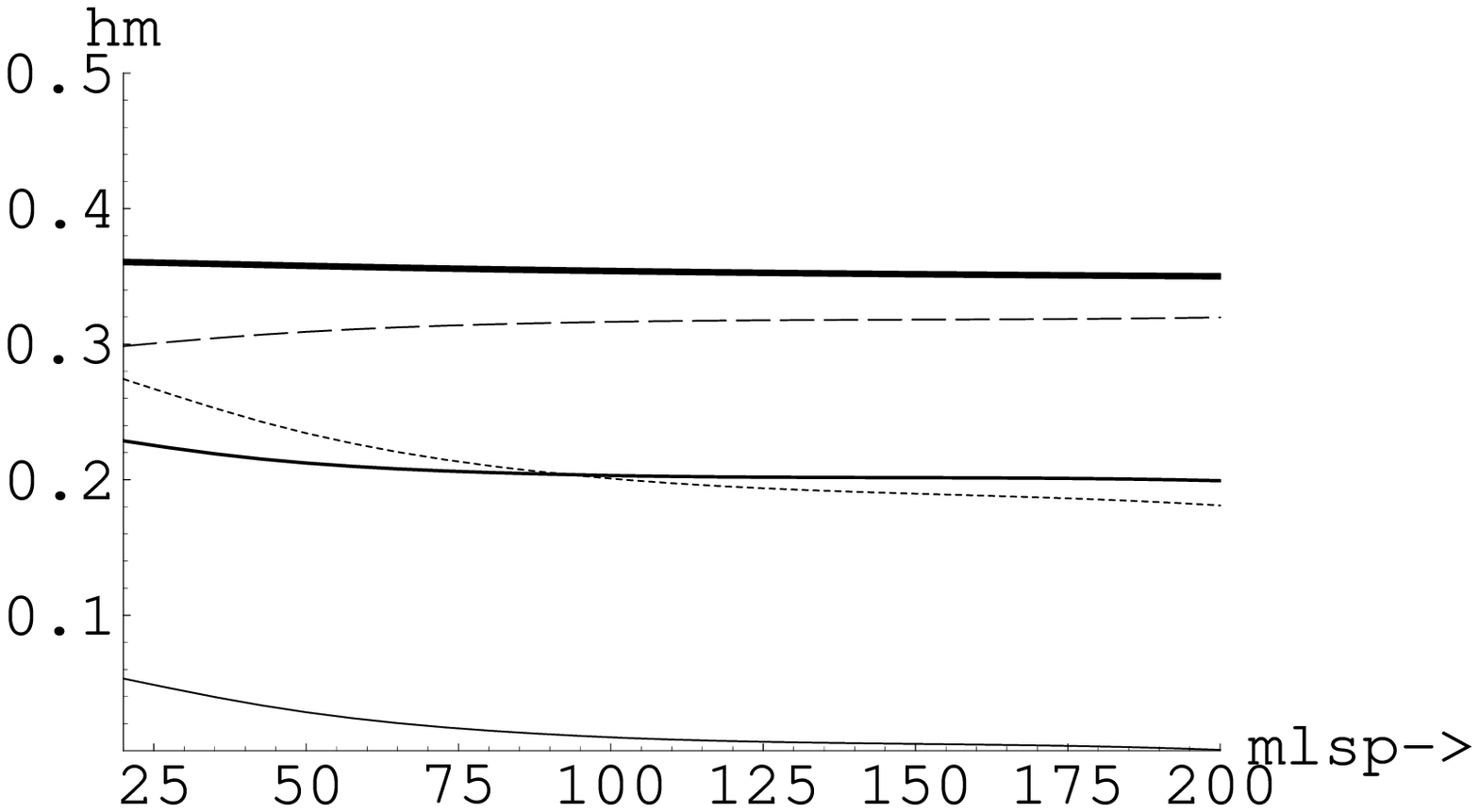}
\caption{ The same as in Fig. \ref{kaph0.0} for $\lambda=1$.
\label{kaph1.0}
}
\end{figure}

 The quantities $\kappa$
and $h_m$ show some variation with the LSP mass (see Figs \ref{kaph0.0}
and \ref{kaph1.0}),
since in this case the reduced mass changes.

\begin{figure}
\includegraphics[height=.3\textheight]{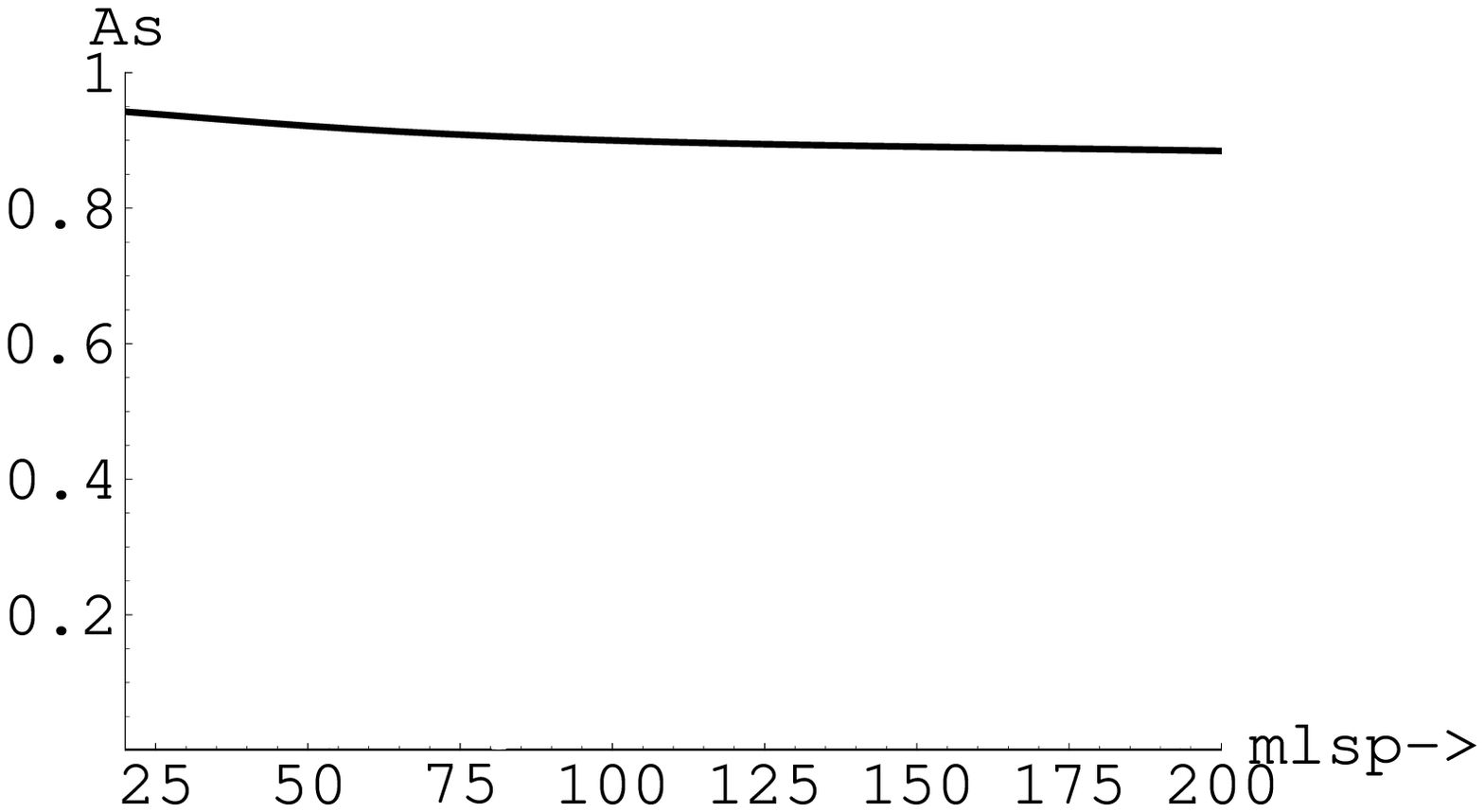}
\includegraphics[height=.3\textheight]{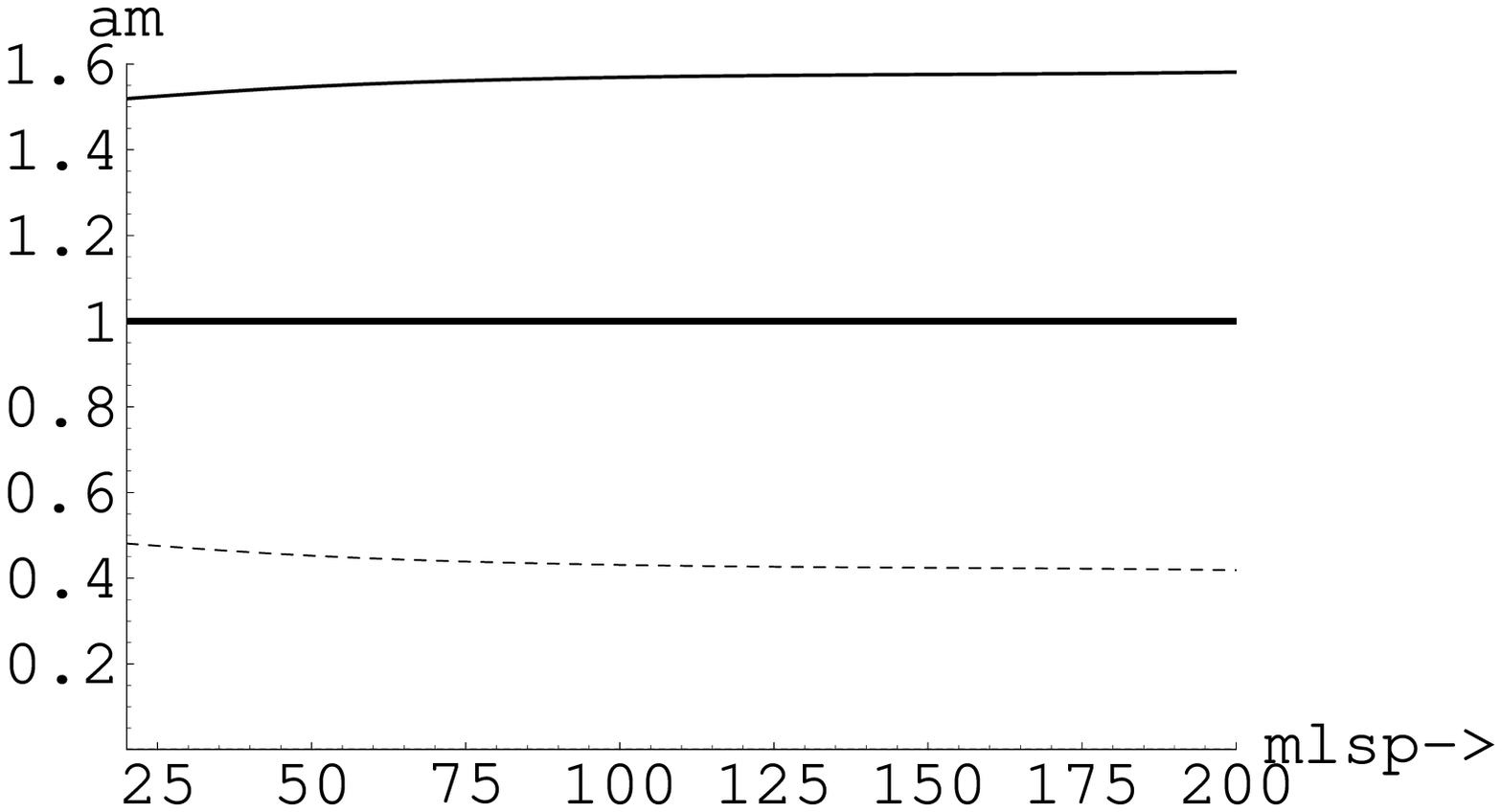}
\caption{ The quantities $As$ and $\alpha_m$ for $\lambda=0$ and
$Q_{min}=0$. The asymmetry $As$, shown on the left, takes only two
values $As \approx 1.0$ in the direction of the sun's motion and
zero in all other directions. These asymmetry plots do not contain
 the contribution due to modulation, since, then, the asymmetry
would depend on the time of observation (see text). On the right
we show the shift (in units of $\pi$) in the position of the
maximum of the modulated amplitude. This shift is almost zero in
the $-z,+y$ directions, close to  $\pi$ in the $+z,-y$ directions,
close to $\pi /2$ in the $+x$ direction and almost $3 \pi /2$ in
the $-x$ direction.
 The notation is the same as in Fig. (\ref{kaph0.0}).
\label{asphi0.0}}.
 \end{figure}

\begin{figure}
\includegraphics[height=.2\textheight]{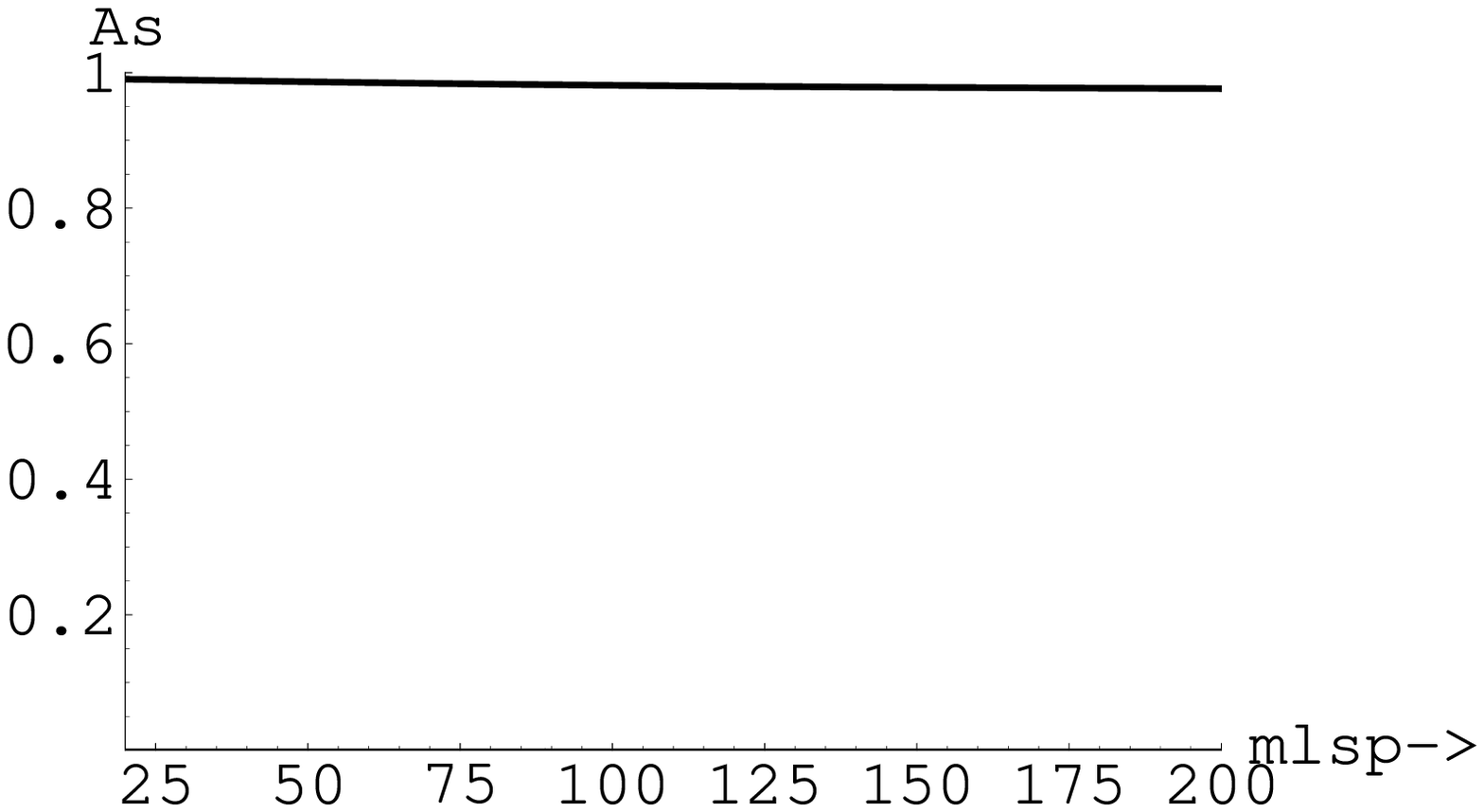}
\includegraphics[height=.2\textheight]{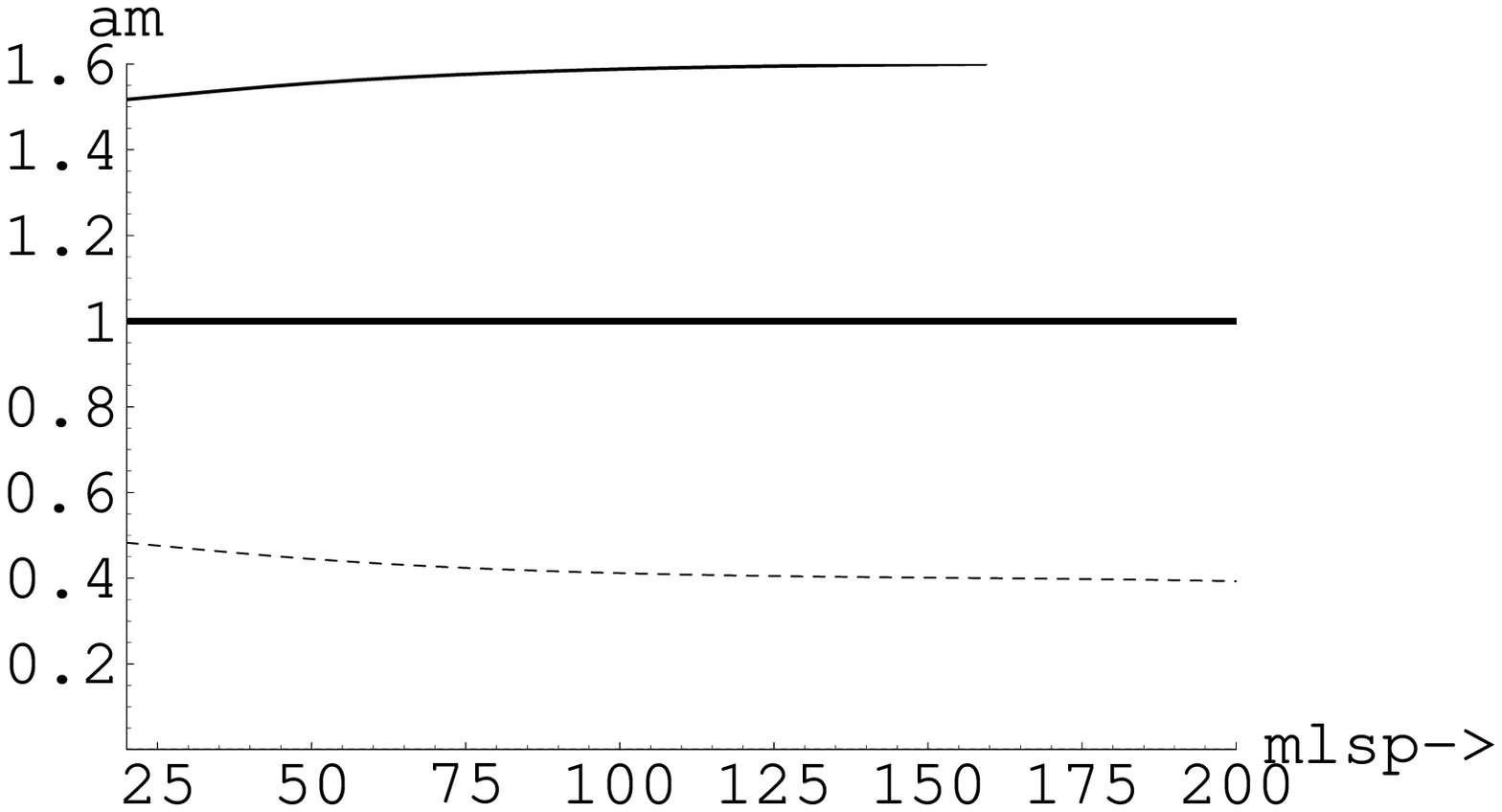}
\caption{ The same as in Fig. \ref{asphi0.0} for $\lambda=1$.
\label{asphi1.0}
}
\end{figure}

 The quantities $As$ and
$\alpha_m$ do not show any significant changes compared to those of the
light systems (see Fig. \ref{asphi0.0} and \ref{asphi1.0}.
We see that, in the absence of modulation, the asymmetry is non zero only
if $\hat{e}$ is in the direction of the sun's motion. In the other directions
the asymmetry depends on the phase of the Earth and is the same with the modulation
, i.e. $h_m|\cos{\alpha}|$ , $h_m|\sin{\alpha}|$ in the y-direction (perpendicular
in the plane of the galaxy) and x-direction (in the radial direction in the
galaxy) respectively.
The results for the light systems are presented in Table \ref{table.dir}.
\subsection{Caustic Rings}
\begin{figure}
\includegraphics[height=.3\textheight]{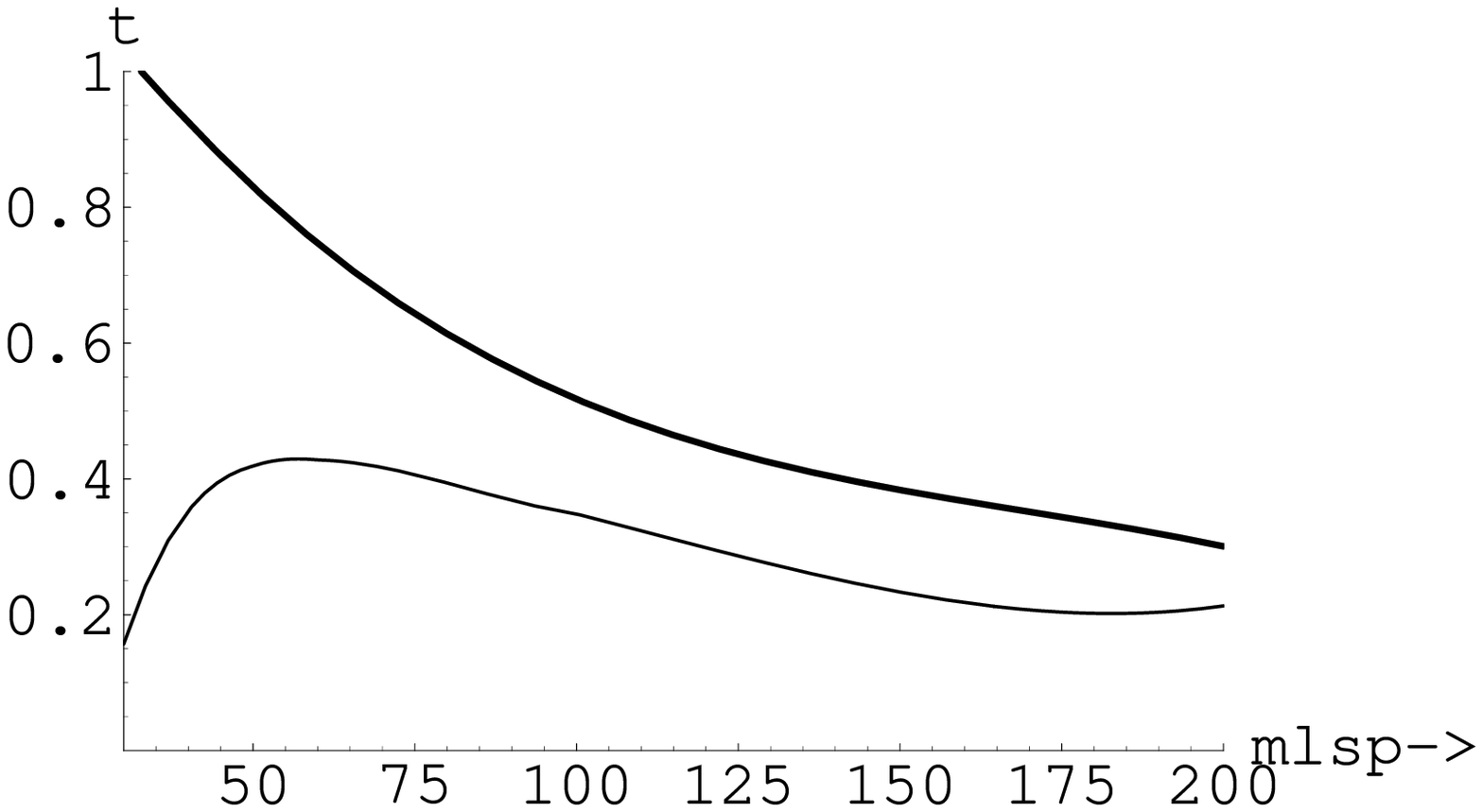}
\includegraphics[height=.3\textheight]{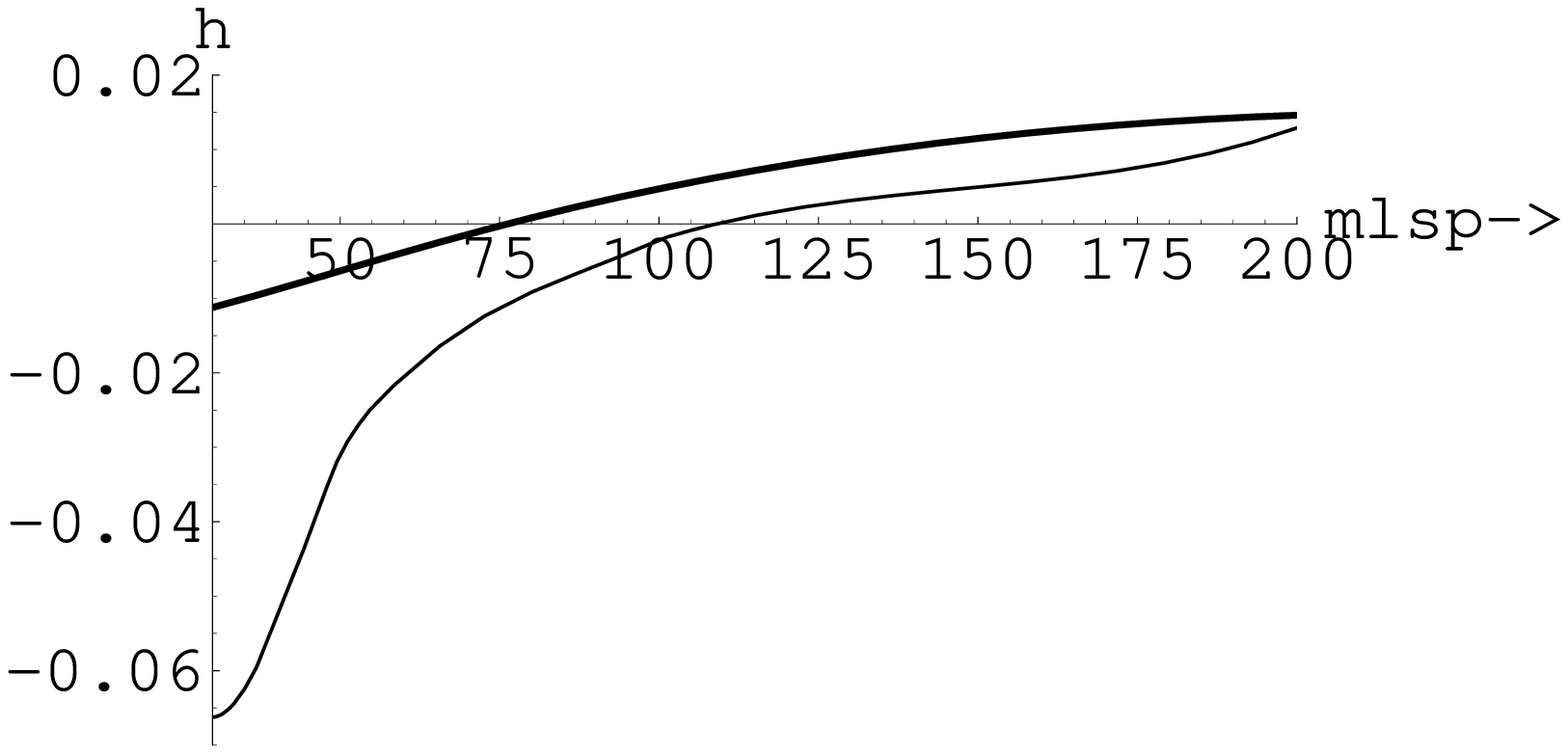}
\caption{ The same as in Fig. \ref{fig.t} (left figure) and \ref{fig.h}
(right figure) in the case of caustic rings.
\label{caus}
}
\end{figure}

\begin{figure}
\includegraphics[height=.2\textheight]{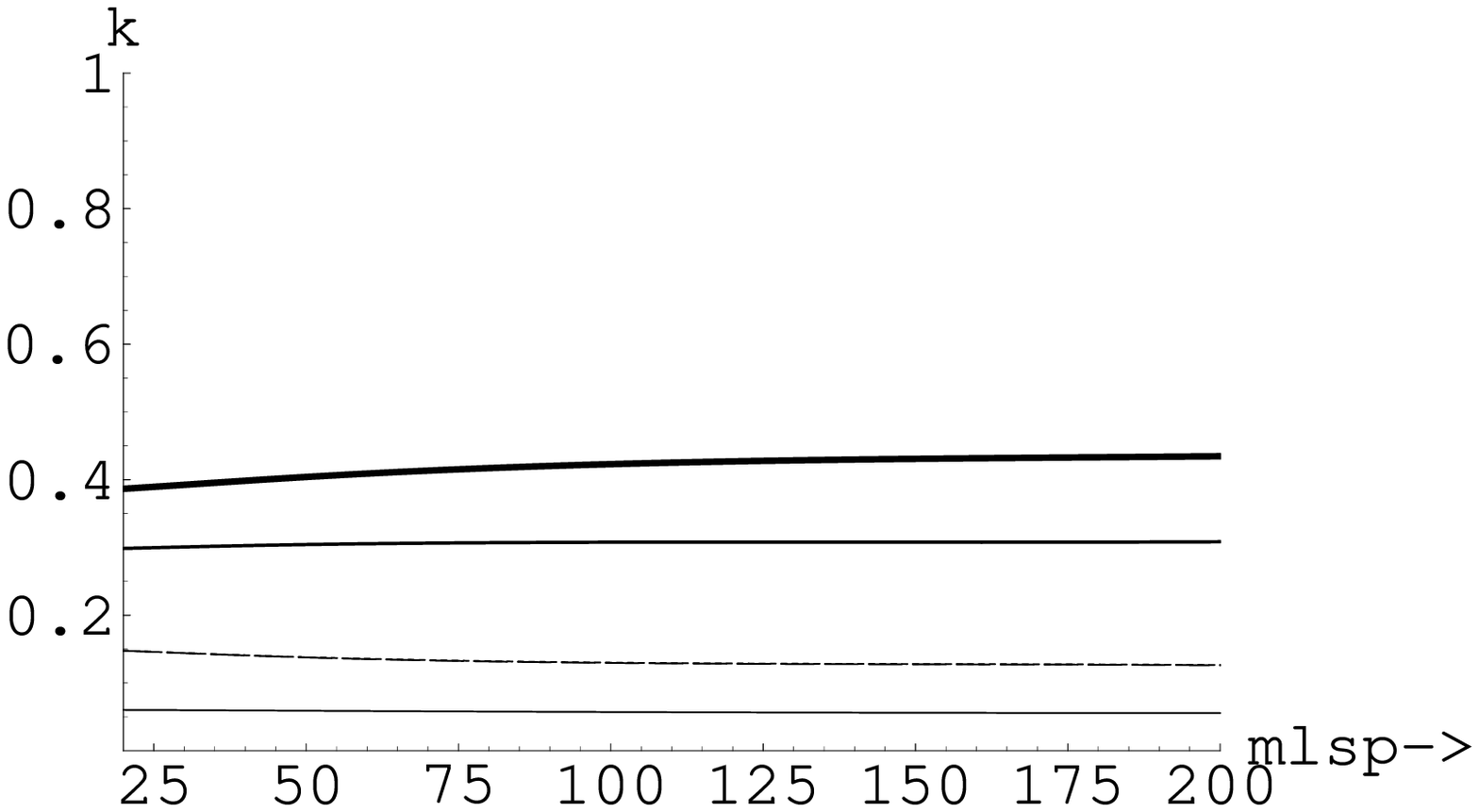}
\includegraphics[height=.2\textheight]{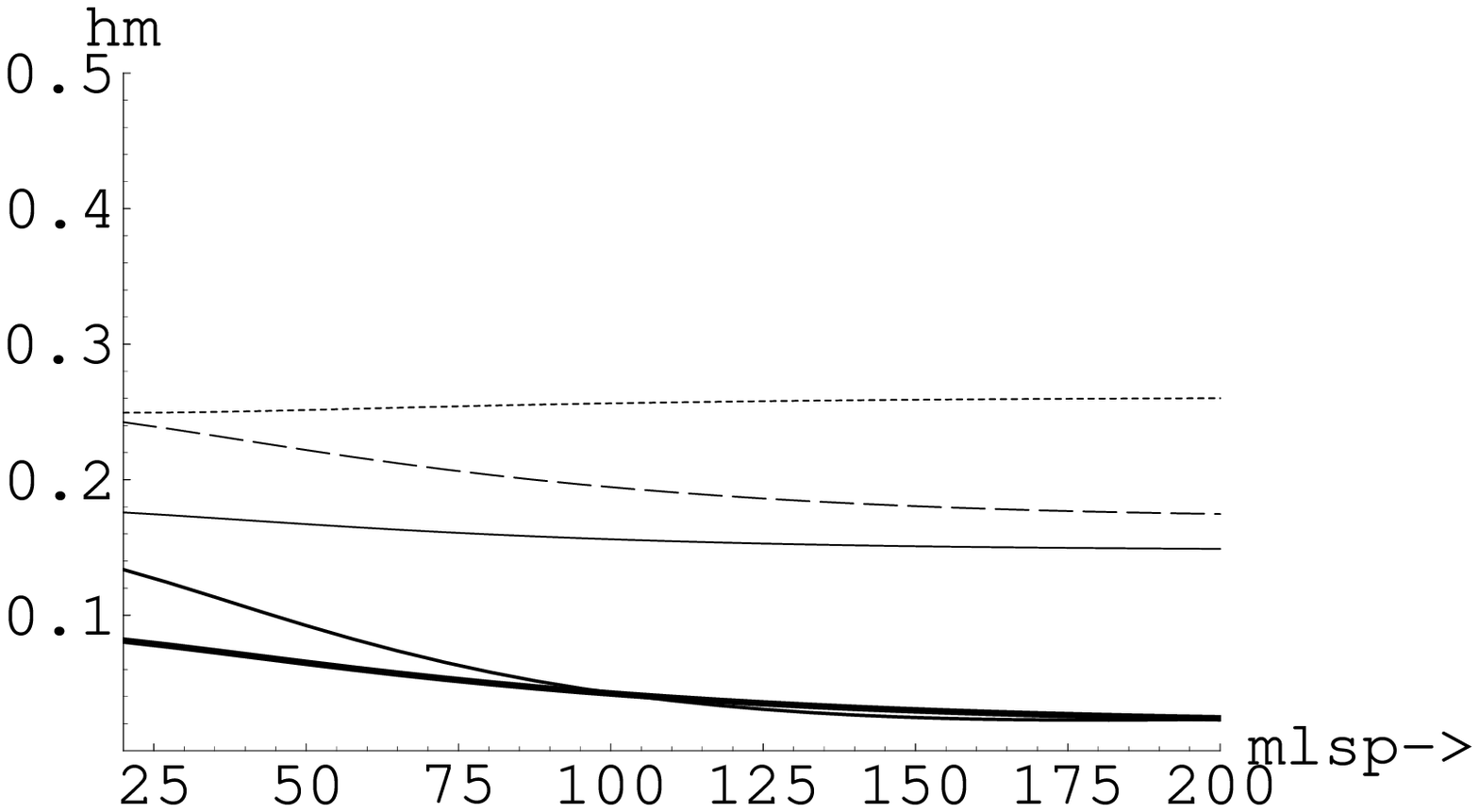}
\caption{ The same as in Fig. \ref{kaph0.0} for caustic rings with $Q_{min}=0$
($\lambda$ is irrelevant).
\label{caus.0}
}
\end{figure}

\begin{figure}
\includegraphics[height=.2\textheight]{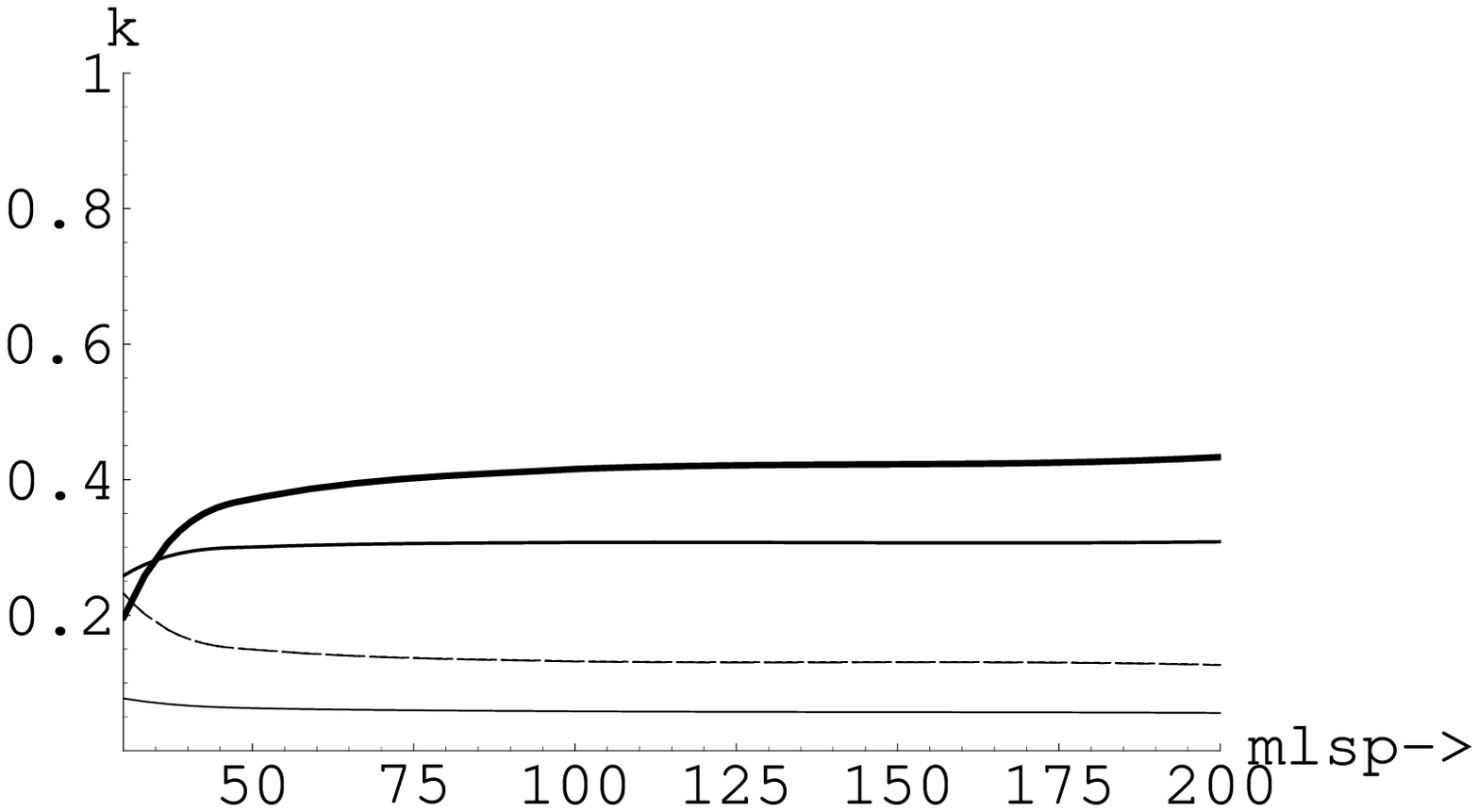}
\includegraphics[height=.2\textheight]{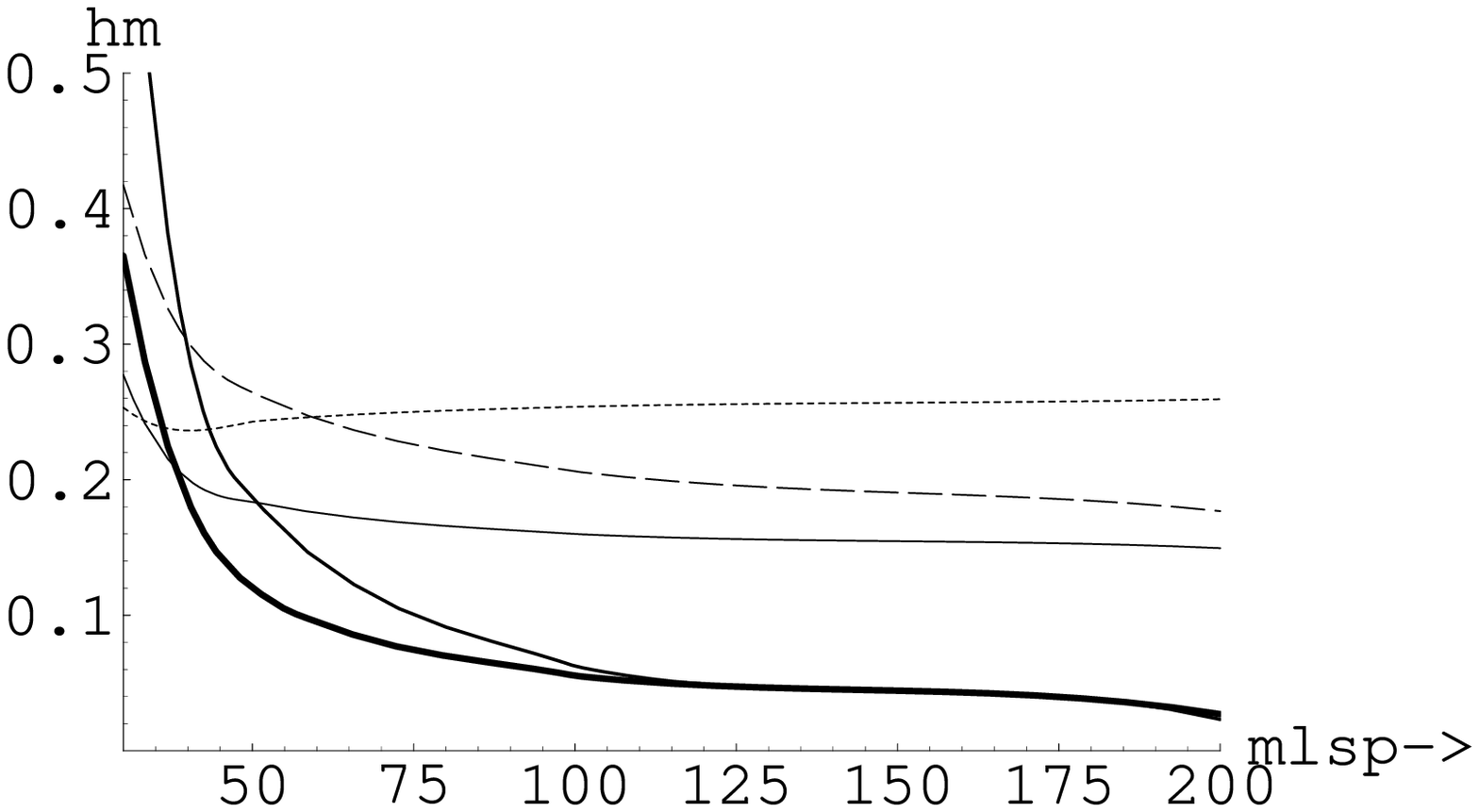}
\caption{ The same as in Fig. \ref{caus.0} for caustic rings with $Q_{min}=10$.
\label{caus.10}
}
\end{figure}

 The case of non isothermal models, e.g. caustic rings, has previously been discussed
\cite{Verg01}in some special cases. Here we will expand our discussion further.
In the case of the non directional rates,
the quantities $t$ and $h$ for $^{127}I$ are shown in Figs \ref{caus}. We see
that as far as $t$ is concerned there is no essential difference between
caustic rings and Gaussian distribution. Notice, however, that the modulation
 is now smaller and of opposite sign.
In the case of directional rates the quantities $\kappa$ and $h_m$ for
 $^{127}I$ are shown in Figs \ref{caus.0}
and \ref{caus.10} for $Q_{min}=0,~10~KeV$ respectively.
The results obtained for the light systems are shown in Table \ref{table.caus2}.
 One clearly sees that
the maximum rate is now in the direction of the sun's motion, $+z$,
 and the minimum
in the $-z$ direction, i.e. in the opposite sense compared with the Gaussian
distribution. In the other directions the rates fall in between. Naturally
the rate is reduced in the presence of an the energy cutoff, but in the case
of caustic rings the reduction manifests itself mainly for small
LSP masses. For such masses it is not easy to have energy above threshold
 transferred to the nucleus.
motion, but smaller than in the Gaussian distribution ($As=0.75$ and $As=0.68$
for $Q_{min}=0$ and 10 $KeV$ respectively). In the other directions the
asymmetry is governed by the modulation.
\section{Conclusions}
In the present paper we have discussed the parameters, which describe
the event rates for direct detection of SUSY dark matter.
Only in a small segment of the allowed parameter space the rates are above
 the present experimental goals ~\cite{Gomez,ref2,ARNDU}.
 We thus looked for
characteristic experimental signatures for background reduction, i.e.
a) Correlation of the event rates
with the motion of the Earth (modulation effect) and b)
 the directional rates (their correlation  both with the velocity of the sun
 and that of the Earth.)

A typical graph for the total non modulated rate is shown Fig. \ref{rate}.
The relative parameters $t$ and $h$ in the case of non directional experiments
are exhibited in Fig. \ref{fig.t} and  Fig. \ref{fig.h} for the Gaussian models.
For caustic rings they are shown in Fig. \ref {caus}.
We must emphasize that the two graphs of Figs. \ref{fig.t} and \ref{caus}
 do not contain
the entire dependence on
the LSP mass. This is due to the fact that there is the extra factor
$m^{-1}_{\chi}$ in Eq. (\ref{3.39b}) and a factor of $\mu_r^2$ arising from
 $\Sigma_i$,
$i=S,~spin,~V$ (see Eqs (\ref{2.10}), (\ref{2.10a}), and  (\ref{2.10d})).
 All these factors combined lead
 to a essentially a constant.
 There remains, however, an LSP mass dependence, which is  due
to the fact that the nucleon cross section itself dramatically depends on the LSP mass.

 Figs \ref{fig.t}, \ref{fig.h}  and \ref{caus} were obtained for
 the scalar interaction, but similar behavior is expected for the spin
 contribution. From the point of view of the static spin matrix elements the
most favored system is the $A=19$ (see our previous work \cite{DIVA00}).
 The scale of the total spin contribution, coming from the SUSY dependent parameter
$\bar{\Sigma}_s$, which was not discussed in this work, may be
very  very different from the corresponding to the scalar
amplitude quantity.
 We should also mention that in the non directional
experiments the modulation $2h$ is small, .i.e. for $\lambda=0$ less than
 $4\%$ for
$Q{min}=0$ and increases to $12\%$ for $Q{min}=10~KeV$ (at the expense of the total
number of counts). For $\lambda=1$ there in no change for
$Q_{min}=0$, but it can go as high as $24\%$ for $Q_{min}=10~KeV$. In the case
of caustic rings is smaller and of opposite sign.

For the directional rates it is instructive to examine the reduction factors
 $\kappa$ if the observation is made in a specific direction, e.g. along the three axes, i.e
along $+z,-z,+y,-y,+x$ and $-x$. These depend on the
nuclear parameters, the reduced mass, the energy cutoff $Q_{min}$ and
 $\lambda$
\cite{Verg00}. Since $f_{red}$ is the ratio of two parameters, its dependence
on $Q_{min}$ and the LSP mass is mild. So we present results for
 $^{127}I$ in Figs
\ref{kaph0.0} and \ref{kaph1.0} (see also Figs \ref{caus.0} and
 \ref{caus.10} for
 caustic rings). In the case of light systems our results are
presented in Tables \ref{table.dir} and \ref{table.caus2}.
 As expected the maximum rate is  along the sun's direction of
 motion, i.e opposite to its velocity $(-z)$ in the Gaussian distribution
 and $+z$ in the case of caustic rings. In fact we find
that $\kappa(-z)$ is around 0.5 ($\lambda=0$) and around 0.6
($\lambda=1.0$). It is not very different from the naively
expected $f_{red}=1/( 2 \pi)~,~\kappa=1$. The asymmetry along the
sun's direction of motion,
$As=|R_{dir}(-)-R_{dir}(+)|/(R_{dir}(-)+R_{dir}(+))$ is quite
characteristic, i.e. almost unity for Gaussian models and a bit
smaller in the case of caustic rings. The rate in the other
directions is quite a bit smaller (see Tables \ref{table.dir} and
\ref{table.caus2}) and the asymmetry is equal to the absolute
value of the modulation.

 The disadvantage of smaller rates in the plane perpendicular to the sun's
 velocity may be compensated by the bonus of very large
and characteristic modulation.

 In conclusion: in the case of directional non modulated rates we expect
 unambiguous correlation with the motion on the sun, which can be explored by
 the experimentalists.
 If one concentrates in a given direction there appears a reduction factor.
 The reduction factor in the most favored direction, .i.e in the line of the
 motion
of the sun, is approximately only $1/(4\pi)$ relative to the non directional
experiments.  In the plane perpendicular to the motion of the sun we expect
interesting modulation signals, but the reduction factor becomes worse.
 These reduction factors do not appear to be an obstacle to the experiments, since
 the TPC counters to be used in the planned experiments can make observations
in almost all directions simultaneously \cite{GIOMATAR}. Once some
candidate events are seen, one can analyze them further in the way
we propose here by selecting those corresponding to a given
direction of observation.

\par
This work was supported by the European Union under the contracts
RTN No HPRN-CT-2000-00148 and TMR
No. ERBFMRX--CT96--0090. During the last stages of this work the author was
in Tuebingen as a Humboldt awardee. He is indebted to professor Faessler for
hospitality and to the Alexander von Humboldt Foundation for its support.

\def\ijmp#1#2#3{{ Int. Jour. Mod. Phys. }{\bf #1~}(#2)~#3}
\def\pl#1#2#3{{ Phys. Lett. }{\bf B#1~}(#2)~#3}
\def\zp#1#2#3{{ Z. Phys. }{\bf C#1~}(#2)~#3}
\def\prl#1#2#3{{ Phys. Rev. Lett. }{\bf #1~}(#2)~#3}
\def\rmp#1#2#3{{ Rev. Mod. Phys. }{\bf #1~}(#2)~#3}
\def\prep#1#2#3{{ Phys. Rep. }{\bf #1~}(#2)~#3}
\def\pr#1#2#3{{ Phys. Rev. }{\bf D#1~}(#2)~#3}
\def\np#1#2#3{{ Nucl. Phys. }{\bf B#1~}(#2)~#3}
\def\npps#1#2#3{{ Nucl. Phys. (Proc. Sup.) }{\bf B#1~}(#2)~#3}
\def\mpl#1#2#3{{ Mod. Phys. Lett. }{\bf #1~}(#2)~#3}
\def\arnps#1#2#3{{ Annu. Rev. Nucl. Part. Sci. }{\bf
#1~}(#2)~#3}
\def\sjnp#1#2#3{{ Sov. J. Nucl. Phys. }{\bf #1~}(#2)~#3}
\def\jetp#1#2#3{{ JETP Lett. }{\bf #1~}(#2)~#3}
\def\app#1#2#3{{ Acta Phys. Polon. }{\bf #1~}(#2)~#3}
\def\rnc#1#2#3{{ Riv. Nuovo Cim. }{\bf #1~}(#2)~#3}
\def\ap#1#2#3{{ Ann. Phys. }{\bf #1~}(#2)~#3}
\def\ptp#1#2#3{{ Prog. Theor. Phys. }{\bf #1~}(#2)~#3}
\def\plb#1#2#3{{ Phys. Lett. }{\bf#1B~}(#2)~#3}
\def\apjl#1#2#3{{ Astrophys. J. Lett. }{\bf #1~}(#2)~#3}
\def\n#1#2#3{{ Nature }{\bf #1~}(#2)~#3}
\def\apj#1#2#3{{ Astrophys. Journal }{\bf #1~}(#2)~#3}
\def\anj#1#2#3{{ Astron. J. }{\bf #1~}(#2)~#3}
\def\mnras#1#2#3{{ MNRAS }{\bf #1~}(#2)~#3}
\def\grg#1#2#3{{ Gen. Rel. Grav. }{\bf #1~}(#2)~#3}
\def\s#1#2#3{{ Science }{\bf #1~}(19#2)~#3}
\def\baas#1#2#3{{ Bull. Am. Astron. Soc. }{\bf #1~}(#2)~#3}
\def\ibid#1#2#3{{ ibid. }{\bf #1~}(19#2)~#3}
\def\cpc#1#2#3{{ Comput. Phys. Commun. }{\bf #1~}(#2)~#3}
\def\astp#1#2#3{{ Astropart. Phys. }{\bf #1~}(#2)~#3}
\def\epj#1#2#3{{ Eur. Phys. J. }{\bf C#1~}(#2)~#3}

\bigskip
\begin{table}[t]
\caption{The quantities
$a_n,yi^{'}_n =\upsilon_n/\upsilon_0,
y_{nz} =\upsilon_{n \phi}/\upsilon_0,
y_{ny} =\upsilon_{nz}/\upsilon_0,
y_{nx} =\upsilon_{nr}/\upsilon_0$ and
$\bar {\rho}_n=d_n/[\sum _{n=1}^{20} d_n]$ and
$y_n=[(y_{nz}-1)^2+y_{ny}^2+y_{nx}^2]^{1/2}$
(for the other definitions see text ).
}
\newpage
\footnotesize
\begin{tabular}{|l|c|rrrrrr|}
\hline
\hline
& & & & & &      \\
n &  $a_{n}(Kpc)$  & $y^{'}_n$  & $y_{nz}$ & $y_{ny}$  & $y_{nx}$
 & $y_n$ & $\bar {\rho}_n $\\
\hline
& & & & & & &      \\
1 &38.0& 2.818& 0.636& $\pm$2.750& 0.000& 2.773 & 0.0120\\
2 &19.0& 2.568& 1.159& $\pm$2.295& 0.000& 2.301 & 0.0301\\
3 &13.0& 2.409& 1.591& $\pm$1.773& 0.000& 1.869 & 0.0601\\
4 & 9.7& 2.273& 2.000& $\pm$1.091& 0.000& 1.480 & 0.1895\\
5 & 7.8& 2.182& 2.000& 0.000& $\pm$0.863& 1.321 & 0.2767\\
6 & 6.5& 2.091& 1.614& 0.000& $\pm$1.341& 1.475 & 0.0872\\
7 & 5.6& 2.023& 1.318& 0.000& $\pm$1.500& 1.533 & 0.0571\\
8 & 4.9& 1.955& 1.136& 0.000& $\pm$1.591& 1.597 & 0.0421\\
9 & 4.4& 1.886& 0.977& 0.000& $\pm$1.614& 1.614 & 0.0331\\
10& 4.0& 1.818& 0.864& 0.000& $\pm$1.614& 1.619 & 0.0300\\
11& 3.6& 1.723& 0.773& 0.000& $\pm$1.614& 1.630 & 0.0271\\
12& 3.3& 1.723& 0.682& 0.000& $\pm$1.591& 1.622 & 0.0241\\
13& 3.1& 1.619& 0.614& 0.000& $\pm$1.568& 1.615 & 0.0211\\
14& 2.9& 1.636& 0.545& 0.000& $\pm$1.545& 1.611 & 0.0180\\
15& 2.7& 1.591& 0.500& 0.000& $\pm$1.500& 1.581 & 0.0180\\
16& 2.5& 1.545& 0.454& 0.000& $\pm$1.477& 1.575 & 0.0165\\
17& 2.4& 1.500& 0.409& 0.000& $\pm$1.454& 1.570 & 0.0150\\
18& 2.2& 1.455& 0.386& 0.000& $\pm$1.409& 1.537 & 0.0150\\
19& 2.1& 1.432& 0.364& 0.000& $\pm$1.386& 1.525 & 0.0135\\
20& 2.0& 1.409& 0.341& 0.000& $\pm$1.364& 1.515 & 0.0135\\
\hline
\hline
\end{tabular}
\label{table.caus}
\end{table}
\begin{table}[t]
\caption{
The quantities $\kappa,h_m,\alpha_m$ and $As$ for the light nuclear systems
(A=19, 23 and 29), are almost identical.
The difference between the coherent
and the spin mode is less than $1\%$. They are also essentially independent
of the LSP mass and the SUSY parameters. Note the phase of the modulation in
the directions $-z~,~+y$ is the same as in the non directional case. The
phase of the modulation in the $-y$ direction is reversed (minimum in June
 3nd). The phase in the $\pm x$ directions leads to a maximum in between.
When the range of a variable is given, it depends somewhat on the LSP mass
(the mass increases to the right).
}
\label{table.dir}
\newpage
\begin{tabular}{lrrrrr}
& & & & &    \\
quantity&dir &$\lambda=0$     &$\lambda=0$   &$\lambda=1.0$  &$\lambda=1.0$ \\
\hline
  &    & $~~~Q_{min}=0.0$&$~~~Q_{min}=10.0$ KeV& $~~~Q_{min}=0.0$& $~~~Q_{min}=10.0$ KeV.\\
\hline
t  &all&1.0  &0.6   &1.0  &0.6        \\
\hline
h  &all&0.02  &0.04-0.06   &0.04  &0.05-0.08        \\
\hline
\footnotesize
            & +z   & 0.018  & 0.010        &  0.003  & 0.000  \\
            &  x   & 0.190  & 0.177        &  0.211  & 0.145 - 0.186  \\
$\kappa$    &  y   & 0.190  & 0.177 - 0.180  &  0.150  & 0.087 - 0.125  \\
            & -z   & 0.690 & 0.758 - 0.760   &  0.752  & 0.519 - 0.672  \\
\hline
            & +z   & 0.211  & 0.242 - 0.226  &  0.361  & 0.380  \\
            & +x   & 0.235  & 0.292 - 0.255  &  0.237  & 0.325 - 0.261  \\
            & +y   & 0.199  & 0.299 - 0.233  &  0.290  & 0.456 - 0.347  \\
$h_m$       & -x   & 0.235  & 0.292 - 0.255  &  0.237  & 0.325 - 0.261  \\
            & -y   & 0.199  & 0.199 - 0.208  &  0.290  & 0.243 - 0.280  \\
            & -z   & 0.060  & 0.100 - 0.068  &  0.063  & 0.158 - 0.092  \\
\hline
            & +z   & 1      & 1            &  1      & 1            \\
            & +x   & 1/2    & 0.445 - 0.484  &  1/2    & 0.432 - 0.467  \\
            & +y   & 0      & 0            &  0      & 0            \\
$\alpha_m$  & -x   & 3/2    & 1.555 - 1.586  &  3/2    & 1.587 - 1.533  \\
            & -y   & 1      & 1            &  1      & 1            \\
            & -z   & 0      & 0            &  0      & 0            \\
\hline
            &  z   & 0.945  & 0.989 - 0.970&  0.991  & 1.000        \\
As        &  x   & $h_m|\sin{\alpha}|$      &  $h_m|\sin{\alpha}|$ &
                   $h_m|\sin{\alpha}|$ & $h_m|\sin{\alpha}|$           \\
As        &  y   & $h_m|\cos{\alpha}|$      &  $h_m|\cos{\alpha}|$ &
                   $h_m|\cos{\alpha}|$ & $h_m|\cos{\alpha}|$           \\
\end{tabular}
\end{table}
\begin{table}[t]
\caption{
The same as in Table \ref{table.dir} in the case of caustic rings for
for the light nuclear systems (A=19, 23 and 29). Note the difference in
the phases between the modulations of the two tables.
}
\label{table.caus2}
\newpage
\begin{tabular}{lrrrrr}
& & & & &    \\
quantity&dir &coherent     &coherent  &spin  &spin \\
\hline
  &    & $~~~Q_{min}=0.0$&$~~~Q_{min}=10.0$ KeV& $~~~Q_{min}=0.0$& $~~~Q_{min}=10.0$ KeV.\\
\hline
t  &all&1.23 - 1.20  &0.547 - 0.893   &1.29 - 1.14  &0.520 - 0.845        \\
\hline
h  &all&-0.015       &-0.057 - (-0.026)   &- 0.015  & -0.057 - (-0.025)     \\
\hline
\footnotesize
            & +z   & 0.381  & 0.310 - 0.364  &  0.383  & 0.310 - 0.363  \\
            &  x   & 0.297  & 0.281 - 0.293  &  0.297  & 0.281 - 0.293  \\
$\kappa$    &  y   & 0.150  & 0.182 - 0.159  &  0.149  & 0.183 - 0.159  \\
            & -z   & 0.060  & 0.067 - 0.062  &  0.060  & 0.068 - 0.062  \\
\hline
            & +z   & 0.086  & 0.208 - 0.112  &  0.083  & 0.310 - 0.113  \\
            & +x   & 0.143  & 0.351 - 0.195  &  0.139  & 0.351 - 0.195  \\
            & +y   & 0.249  & 0.239 - 0.246  &  0.249  & 0.239 - 0.246  \\
$h_m$       & -x   & 0.144  & 0.351 - 0.195  &  0.137  & 0.351 - 0.195  \\
            & -y   & 0.247  & 0.315 - 0.267  &  0.245  & 0.281 - 0.267  \\
            & -z   & 0.178  & 0.209 - 0.187  &  0.178  & 0.209 - 0.187  \\
\hline
            & +z   & 1      & 1            &  1      & 1            \\
            & +x   & 3/2    & 3/2          &  3/2    & 3/2  \\
            & +y   & 0      & 0            &  0      & 0            \\
$\alpha_m$  & -x   & 1/2    & 1/2          &  1/2    & 1/2  \\
            & -y   & 1      & 1            &  1      & 1            \\
            & -z   & 0      & 0            &  0      & 0            \\
\hline
          &  z   & 0.726  & 0.667        &  0.728    & 0.667        \\
As        &  x   & $h_m|\sin{\alpha}|$      &  $h_m|\sin{\alpha}|$ &
                   $h_m|\sin{\alpha}|$ & $h_m|\sin{\alpha}|$           \\
As        &  y   & $h_m|\cos{\alpha}|$      &  $h_m|\cos{\alpha}|$ &
                   $h_m|\cos{\alpha}|$ & $h_m|\cos{\alpha}|$           \\
\end{tabular}
\end{table}

\begin{thebibliography}{8.}
\addcontentsline{toc}{section}{References}
\bibitem{MAXIMA-1}
S. Hanary {\it et al}, {\it Astrophys. J.} {\bf 545}, L5 (2000);\\
J.H.P Wu {\it et al}, {\it Phys. Rev. Lett.} {\bf 87}, 251303 (2001);\\
M.G. Santos {\it et al}, {\it Phys. Rev. Lett.} {\bf 88}, 241302
(2002)
\bibitem{BOOMERANG}
P.D. Mauskopf {\it et al}, {\it Astrophys. J.} {\bf 536}, L59 (20002);\\
S. Mosi {\it et al}, {\it Prog. Nuc.Part. Phys.} {\bf 48}, 243 (2002);\\
S.B. Ruhl {\it al}, astro-ph/0212229 and references therein.
\bibitem{DASI}
N.W. Halverson {\it et al}, {Astrophys. J.} {\bf 568}, 38 (2002)\\
L.S. Sievers {\it et al},astro-ph/0205287 and references therin.
\bibitem{COBE} G.F. Smoot {\it et al}, (COBE data), {\it Astrophys. J.} {\bf
396}, (1992) L1.
\bibitem{flat01}
A.H. Jaffe {\it et al.},{\it Phys. Rev. Lett.} {\bf 86}, 3475
(2001).
\bibitem{GAW}E. Gawiser and J. Silk,{\it {Science}} {\bf 280}, 1405 (1988);
 M.A.K Gross, R.S. Somerville, J.R.  Primack, J. Holtzman  and  A.A. Klypin,
  {\it Mon. Not. R. Astron. Soc.} {\bf 301}, 81 (1998).
\bibitem{HSST} A.G. Riess {\it et al}, {\it Astron. J.}
{\bf 116} , 1009 (1998).
\bibitem{SPF} R.S. Somerville, J.R. Primack and S.M. Faber, astro-ph/9806228;
{\it Mon. Not. R. Astron. Soc.} (in press).
\bibitem{SCP}S. Perlmutter {\it et al} {\it Astrophys. J.}
 {\bf 517}, 565 (1997); {\bf 483}, 565 (1999); {\it astro-ph}/9812133.\\
S. Perlmutter,  M.S. Turner and M. White, {\it Phys. Rev. Let.} {\bf 83},
670 (1999).
\bibitem{Primack} J.R. Primack, astro-ph/0205391
\bibitem{Eina01} Jaan Einasto, in Dark Matter in Astro- and Particle Physics,
p.3, Ed. H.V. Klapdor-Kleingrothaus, Springer-Verlag Berlin Heidelberg 2001.
\bibitem{Benne} D.P. Bennett {\it et al.}, (MACHO collaboration), A binary
lensing event toward the LMC: Observations and Dark Matter Implications,
Proc. 5th Annual Maryland Conference, edited by S. Holt (1995);\\
C. Alcock {\it et al.}, (MACHO collaboration), {\it Phys. Rev. Lett.} {\bf 74}
, 2967 (1995).
\bibitem{BERNA2} R. Bernabei et al., INFN/AE-98/34, (1998);
 R. Bernabei et al., {it Phys. Lett.} {\bf B 389}, 757 (1996).
\bibitem{BERNA1} R. Bernabei et al., {\it Phys. Lett.} {\bf B 424}, 195 (1998);
{\bf B 450}, 448 (1999).
\bibitem{GOODWIT}
M.W. Goodman and E. Witten, {\it  Phys. Rev. D} {\bf 31}, 3059
(1985).
\bibitem{GRIEST}
K. Griest, {\it Phys. Rev. Lett} {\bf  61}, 666 (1988).
\bibitem{ELLFOR}
J. Ellis, and R.A. Flores, {\it Phys. Lett. B} {\bf 263}, 259
(1991); {\it Phys. Lett. B} {\bf 300}, 175 (1993); {\it Nucl.
Phys. B} {\bf 400}, 25 (1993).
\bibitem{ELLROSZ}
J. Ellis and L. Roszkowski, {\it Phys. Lett. B} {\bf  283}, 252
(1992).
\bibitem{ref1}For more references see e.g. our previous report:\\
J.D. Vergados,  Supersymmetric Dark Matter Detection-
The Directional Rate and the Modulation Effect, hep-ph/0010151;
\bibitem{Gomez}  M.E.G\'{o}mez and  J.D. Vergados, {\it Phys. Lett. B} {\bf 512}
, 252 (2001); hep-ph/0012020.\\
 M.E. G\'{o}mez,  G. Lazarides and Pallis, C.,\pr{61}{2000}{123512} and
 {\it Phys. Lett. B} {\bf 487}, 313 (2000).
\bibitem{gtalk}  M.E. G\'{o}mez and J.D. Vergados, hep-ph/0105115.
\bibitem{ref2}
A. Bottino {\it et al.}, {\it Phys. Lett B} {\bf  402}, 113 (1997).\\
R. Arnowitt. and P. Nath, {\it Phys. Rev. Lett.}  {\bf 74}, 4952 (1995);
 {\it Phys. Rev. D} {\bf 54}, 2394 (1996); hep-ph/9902237;\\
 V.A. Bednyakov,  H.V. Klapdor-Kleingrothaus and  S.G. Kovalenko,
{\it Phys. Lett. B}  {\bf  329}, 5 (1994).
\bibitem{JDV96} J.D. Vergados, {\it J. of Phys. G} {\bf  22}, 253 (1996).
\bibitem{ARNDU}
 A. Arnowitt and B. Dutta, Supersymmetry and Dark Matter,
hep-ph/0204187.
\bibitem{AADS00}
 E. Accomando, A. Arnowitt and B. Dutta, Dark Matter, muon G-2 and other
 accelerator constaints, hep-ph/0211417.
\bibitem{KVprd}T.S. Kosmas and  J.D. Vergados, {\it Phys. Rev. D} {\bf 55},
1752 (1997).
\bibitem{drees} M. Drees and N,M. Nojiri,
\pr{47}{1993}{376};
 (1985).
\bibitem{Dree}M. Drees and  N.N. Nojiri, {\it Phys. Rev. D} {\bf 48},
3843 (1993); {\it Phys. Rev. D} {\bf  47}, 4226 (1993).
\bibitem{Dree00}A. Djouadi and MK. Drees, {\it Phys. Lett. B} {\bf 484}, 183 (2000);
S. Dawson, {\it Nucl. Phys.} {\bf B359}, 283 (1991); M. Spira {it
et al}, {\it Nucl. Phys.} {\bf B453}, 17 (1995).
\bibitem{Chen} T.P. Cheng, {\it Phys. Rev. D} {\bf  38}, 2869 (1988);
 H-Y. Cheng, {\it Phys. Lett. B} {\bf  219}, 347 (1989).
\bibitem{Ress} M.T. Ressell {\it et al.}, {\it Phys. Rev. D} {\bf  48},
 5519 (1993);
\bibitem{KVdubna} J.D. Vergados and  T.S. Kosmas,  {\it Physics of Atomic
 nuclei}, Vol. {\bf 61}, No 7, 1066 (1998)
(from {\it Yadernaya Fisika}, Vol. 61, No 7, 1166 (1998).
\bibitem{DIVA00} P.C. Divari, T.S. Kosmas, J.D. Vergados and L.D. Skouras,
{\it  Phys. Rev. C} {\bf 61} (2000), 044612-1.
\bibitem{DFS86}
A.K. Drukier, K. Freese and D.N. Spergel, {\it Phys. Rev. D} {\bf
33}, 3495 (1986).
\bibitem{FFG88}
K. Frese, J.A Friedman, and A. Gould, {\it Phys. Rev. D} {\bf 37},
3388 (1988).
\bibitem{Verg98}J.D. Vergados, {\it Phys. Rev. D} {\bf  58}, 103001-1 (1998).
\bibitem{Verg99}J.D. Vergados, {\it Phys. Rev. Lett} {\bf  83}, 3597
(1999).
\bibitem{Verg00}J.D. Vergados, {\it Phys. Rev. D} {\bf  62}, 023519 (2000).
\bibitem{Verg01} J.D. Vergados, {\it Phys. Rev. D} {\bf 63}, 06351 (2001).
\bibitem{COLLAR92}
J.I. Collar {\it et al}, {\it Phys. Lett. B} {\bf 275}, 181
(1992).
\bibitem{UK01}
P. Ullio and M. Kamioknowski, {\it JHEP} {\bf 0103}, 049 (2001).
\bibitem{BCFS02}
P. Belli, R. Cerulli, N. Fornego and S. Scopel, {\t Phys. Rev. D}
{\bf 66}, 043503 (2002); hep-ph/0203242.
\bibitem{GREEN02}
A. Green, {\it Phys. Rev. D} {\bf 66}, 083003 (2002)
\bibitem{UKDMC}K.N. Buckland, M.J. Lehner and  G.E. Masek, in
 Proc. {\it 3nd Int. Conf. on Dark Matter
in Astro- and part. Phys.} (Dark2000), Ed. H.V. Klapdor-Kleingrothaus,
Springer Verlag (2000).
\bibitem{VALLS} {\it CDF Collaboration}, FERMILAB-Conf-99/263-E CDF;\\
http://fnalpubs.fnal.gov/archive/1999/conf/Conf-99-263-E.html.
\bibitem{DORMAN}
 P.J. Dorman {\it ALEPH Collaboration} March 2000,\\
http://alephwww.cern.ch/ALPUB/seminar/lepc$_{}$mar200/lepc2000.pdf.
\bibitem{LEP2}
L3 collabtation, M. Acciari {\it et al}, {\it Phys. Lett. B}
{\bf 495}, 18 (2000); hep-ex/0011043\\
ALEPH collabtation, R. Barate {\it et al}, {\it Phys. Lett. B}
{\bf 495}, 1 (2000); hep-ex/0011045\\
DELPHI collabtation, P. Abreu {\it et al}, {\it Phys. Lett. B}
{\bf 499}, 23 (2001)\\
OPAL collabtation, G. Abbiendi {\it et al}, {\it Phys. Lett. B}
{\bf 499}, 38 (2000).
\bibitem{BFS01}
A. Bottino, N. Fornengo and S. Scopel, {\it Nucl. Phys. B} {\bf
608}, 461 (2001); hep-ph/0212379.
\bibitem{ADHSW}
S. Ambrosanio, A. Dedes, S.Heinemeyer, S. Su, and G. Weiglein,
{\it Nuc. Phys. B} {\bf 624}, 3 (2002); hepph/0106255.
\bibitem{JDV02}
J.D. Vergados, SUSY Dark Matter in Universe- Theoretical Direct
Detection Rates,
 Proc. {\it NANP-01, International Conference on Non Accelerator New Physics},
Dubna, Russia, June 19-23, 2001, Editors V. Bednyakov and S. Kovalenko,
hep-ph/0201014.
\bibitem{BDFS99}
A. Bottino, F. Donato, N. Forengo and S. Scopel {\it Phys. Rev. D}
{\bf 59}, 095004 (1999).
\bibitem{AAD00}
E. Accomendo, R. Arnowitt, B. Dutta and Y. Santoso, {\it Nuc.
Phys. B} {\bf 585}, 124 (2000).
\bibitem{EDDIN}
A.S. Eddington {\it NRAS} {\bf 76}, 572 (1916);\\
D. Merrit, A J {\bf 90}, (1985)
\bibitem{VEROWEN}
J.D. Vergados and D. Owen, to appear in ApJ, astro-ph/0203293
\bibitem{SIKIVIE}P. Sikivie, I. Tkachev and Y. Wang, {\it Phys. Rev. Let.} {\bf 75},
2911 (1995; {\it Phys. Rev. D} {\bf 56}, 1863 (1997); {\it Phys. Rev. D} {\bf 60}, 063501 (1999)\\
P.Sikivie, {\it Phys. Lett. B} {\bf  432}, 139 (1998);
 astro-ph/0109296,astro-ph/9810286
\bibitem{GEL01} G. Gelmini and P. Gondolo, {\it Phys. Rev. D} {\bf 64}, 023504 (2001)
\bibitem{MHWS}
B. Moore {\it et al}, {\it Phys. Rev. D} {\bf 64}, 063508 (2001);\\
B. Moore, {\t IDM 2000, 3nd Workshop on the Ident. of Dark
Matter}, Ed.
N. Spooner, World Scientific, astro-ph/0103094\\
A. Helmi, S.D.M. White and V. Springer, {\it Phys. Rev. D} {\bf
66}, 063502, astro-ph/0201289
\bibitem{SIMON03}
E. Simon {\it et al}, SICANE:A detector Array for the Measurement
of Nuclear Recoil Quenching Factors using a Monoenergetic Neutron
Beam, astro-ph/0212491.
\bibitem{GRAICHEN}
J. Graichen {\it et al}, {\it Nucl. Instr. Meth. A} {\bf 485}, 774
(2002).
\bibitem{GERBIER}
G. Gerbier {\it et al}, {\bf Astropart. Phys. } {\bf 11}, 287 (1999);\\
D.R. Tovey {\it et al}, {\it Phys. Lett. B} {\bf 433}, 150 (1998);\\
J.J.C. Spooner {\it et al}, {\it Phys. Lett. B} {\bf 321}, 156
(1994).
\bibitem{GIOMATAR}
Y. Giomataris, Ph. Rebourgeant, J.P. Robert and C. Charpak,
 {\it Nucl. Instr. Meth A} {bf 376}, 29 (1996);\\
J.I. Collar, Y. Giomataris, Low background applications of
Micromegas detector technology, Presented at {\it IMAGING 2000},
Stockholm 28.06.2000-01.07.2000, DAPNIA/00-08;\\
J. Bouchez and Y. Giomataris, Private communication.
\end{thebibliography}
\end{document}